\documentclass[onecolumn,showpacs,nobibnotes,nofootinbib,12pt,aps,prd,showpacs,notitlepage,nofootinbib,preprintnumbers,amsmath,amssymb]{article}
\pdfoutput=1

\usepackage{jheppub}

\usepackage{amsmath,amssymb}
\usepackage{wasysym}
\usepackage{slashed}
\usepackage{graphicx}
\usepackage{pdfcomment}

\usepackage{graphics,graphicx}
\usepackage{dcolumn}
\usepackage{bm}
\usepackage{epsfig}
\usepackage{hyperref} 
\usepackage{mathbbol}
\usepackage{epstopdf}
\usepackage{simplewick}
\usepackage[utf8]{inputenc} 
\usepackage{slashed}

\def\eq#1{{Eq.~(\ref{#1})}}
\def\fig#1{{Fig.~\ref{#1}}}
\newcommand{\ben}{\begin{eqnarray*}}
\newcommand{\een}{\end{eqnarray*}}
\newcommand{\un}[1]{\underline{#1}}

\newcommand{\as}{\alpha_s}

\newcommand{\dhd}{{\textstyle d}
\lower.03ex\hbox{\kern-0.38em$^{\scriptstyle-}$}\kern-0.05em{}}
\newcommand{\dbar}{{\textstyle \delta}
\lower.03ex\hbox{\kern-0.38em$^{\scriptstyle-}$}\kern-0.05em{}}

\newcommand{\Tr}{\mathrm{Tr}}

\title{Time-Dependent Observables in Heavy Ion Collisions I: Setting
  up the Formalism}

\author{Bin Wu,}
\author{Yuri V. Kovchegov}

\affiliation{Department of Physics, The Ohio State University,
  Columbus, OH 43210, USA}

\emailAdd{bin.wu.phys@gmail.com}
\emailAdd{kovchegov.1@osu.edu}

\abstract{We adapt the Schwinger-Keldysh formalism to study heavy-ion
  collisions in perturbative QCD. Employing the formalism, we
  calculate the two-point gluon correlation function
  $G_{22}^{a\mu,b\nu}$ due to the lowest-order classical gluon fields
  in the McLerran-Venugopalan model of heavy ion collisions and
  observe an interesting transition from the classical fields to the
  quasi-particle picture at later times. Motivated by this
  observation, we push the formalism to higher orders in the coupling
  and calculate the contribution to $G_{22}^{a\mu,b\nu}$ coming from
  the diagrams representing a single rescattering between two of the
  produced gluons. We assume that the two gluons go on mass shell both
  before and after the rescattering. The result of our calculation
  depends on the ordering between the proper time of the rescattering
  $\tau_Z$ and the proper time $\tau$ when the gluon distribution is
  measured. For (i) $\tau_Z\gg 1/Q_s$ and $\tau-\tau_Z\gg 1/Q_s$ (with
  $Q_s$ the saturation scale) we obtain the same results as from the
  Boltzmann equation. For (ii) $\tau-\tau_Z\gg \tau_Z\gg 1/Q_s$ we end
  up with a result very different from kinetic theory and consistent
  with a picture of ``free-streaming'' particles. Due to the
  approximations made, our calculation is too coarse to indicate
  whether the ordering (i) or (ii) is the correct one: to resolve this
  controversy, we shall present a detailed diagrammatic calculation of
  the rescattering correction in the $\varphi^4$ theory in the second
  paper of this duplex.}

\begin{document}

\maketitle
\flushbottom


\section{Introduction}

The ultimate goal of heavy-ion collision is to produce and study
quark-gluon plasma (QGP). QGP is believed to be the primordial matter
in our early universe after the Big bang and before the formation of
nucleons. Heavy-ion collision experiments at RHIC and LHC provide us a
golden opportunity to study such a new form of matter. The detectors
only measure the properties of the system at a very late time
$\tau\sim 10^{15}~\text{fm}/c$ after the collision. QGP is believed to
exist only in the first 10-20~fm/$c$ after the collision: for the
majority of the remaining time the system reduces to a multitude of
hadrons free-streaming towards the detector. The entire evolution
history of bulk matter in the first 10-20~fm/$c$ can be studied only
by comparing theoretical calculations to the experimental data at
roughly $10^{15}~\text{fm}/c$. However, a consistent first-principles
QCD formalism allowing to study the system from the very beginning of
the collision to a moderate late time is still missing.

Hydrodynamic models could give a good description of the collective
behavior seen in the experimental data (see \cite{Heinz:2013th} for a
recent review). Hydrodynamics can be taken as an effective theory of
the underlying quantum field theory near (local) thermal equilibrium
\cite{Jeon:1995zm}. Since bulk matter in heavy-ion collisions is far
from thermal equilibrium at the very early stage, hydrodynamics breaks
down at those early times. In practice, hydrodynamic models are
switched on at some initial time $\tau_0$ with the initial condition
provided by other theoretical studies. In order to study the early
stage of the collision one needs to employ the underlying field
theory.

At the very early stage of a collision, a large number of saturated
gluons are believed to be freed from the two nuclear wave functions
(see \cite{Gelis:2010nm, KovchegovLevin} for a comprehensive
review). In this case the classical Yang-Mills theory applies. It has
been extensively studied in
\cite{McLerran:1993ni,McLerran:1993ka,McLerran:1994vd,Kovchegov:1996ty,Ayala:1996hx,Kovchegov:1997ke,Krasnitz:1998ns,Krasnitz:1999wc,Krasnitz:2003nv,Lappi:2003bi,Kovchegov:2005ss,Berges:2013fga,Gelis:2013rba}. However,
this approach can only give a highly anisotropized energy-momentum
tensor with the ratio of the longitudinal to transverse pressures
$P_L/P_T$ approaching zero at later times
\cite{Krasnitz:1998ns,Krasnitz:1999wc,Krasnitz:2003nv,Kovchegov:2005ss,Berges:2013fga}. Early
pressure isotropization has been observed if certain types of vacuum
quantum fluctuations are included in the classical field simulation
\cite{Epelbaum:2013waa,Gelis:2013rba}. In this case one has to deal with the
dependence of the medium energy-momentum tensor on the lattice spacing
\cite{Gelis:2013rba,Berges:2013lsa}. This is a consequence of the
non-renormalizability of the classical field approach with vacuum
quantum fluctuations \cite{Epelbaum:2014yja,Epelbaum:2014mfa}.

The Boltzmann equation has also been broadly used in heavy-ion
collisions. It can be derived from two-point Green functions in
quantum theory using the so-called quasi-particle approximation near
thermal equilibrium
\cite{Kadanoff,Chou:1984es,Calzetta:1986cq,Blaizot:2001nr,Arnold:2002zm}. The
transition from classical fields to quasi-particles is expected to
occur at $\tau\sim 1/Q_s$ with $Q_s$ the saturation momentum of the
colliding nuclei \cite{Baier:2000sb}. Then, a parametric estimate
using the quasi-particle picture gives a bottom-up scenario for the
system to establish thermal equilibrium \cite{Baier:2000sb}. This
picture has recently been confirmed by numerical solutions of the
Boltzmann equation \cite{Kurkela:2015qoa}. One of the intriguing
questions about the Boltzmann equation is when it starts to apply to
heavy-ion collisions since the derivation of this equation from
quantum field theory has mostly been done for the systems near thermal
equilibrium. The conventional understanding is that when the gluon
density $f$ is less than $1/g^2$ (with $g$ the QCD coupling), both the
Boltzmann equation and classical field approximation apply
\cite{Mueller:2002gd}. However, this argument is based on the
so-called quasi-particle approximation. It is of great interest to
understand whether and how such a transition occurs in the collision
process.

The Schwinger-Keldysh formalism or the close-time path formalism was
first invented by Schwinger \cite{Schwinger:1960qe} and Keldysh
\cite{Keldysh:1964ud}. It gives a unified description of equilibrium
and non-equilibrium systems in quantum field theory
\cite{Chou:1984es}. This formalism has been used to study thermal
equilibrium systems in thermal field theory
\cite{Landsman:1986uw,Bellac:2011kqa}. It has also been used to study
non-equilibrium systems by resumming two-particle-irreducible (2PI) or
n-particle-irreducible (nPI) diagrams
\cite{Cornwall:1974vz,Calzetta:1986cq,Berges:2004pu}. The interested
reader is referred to \cite{Berges:2004yj,Calzetta:1986cq} for a
comprehensive review for the nPI effective theories. A 2PI non-Abelian
gauge theory would be of great interest to heavy-ion physics. However,
the truncated 2PI effective action leads to gauge-dependent results
for most observables \cite{Carrington:2003ut}. In high-energy nuclear
physics, the Schwinger-Keldysh formalism has been employed to resum
leading $\ln \frac{1}{x}$ terms with $x$ the energy fraction into the
color charge density functionals describing the colliding nuclei
\cite{Gelis:2008rw,Gelis:2008ad,Jeon:2013zga}. However, contributions
beyond the leading $\ln \frac{1}{x}$ have not been evaluated: such
contributions could be important for the evolution of the system at
late times.

The main purpose of this paper is to adapt the Schwinger-Keldysh
formalism to study heavy-ion collisions in a perturbative approach.
This approach is obviously gauge invariant. This paper is organized as
follows. We first give a brief review of this formalism and derive the
Feynman rules for perturbative calculations in Sec. \ref{sec:sk}.  In
Sec. \ref{sec:quasi} we recalculate the gluon two-point function by
using the lowest-order classical gluon fields of the
McLerran-Venugopalan (MV) model
\cite{McLerran:1993ni,McLerran:1993ka,McLerran:1994vd} in the
light-cone gauge.  Based on this calculation, we show explicitly how
quasi-particles emerge from classical fields. In Sec.~\ref{sec:resc}
we study the contribution from a $2\to 2$ rescattering of these
quasi-particles to the two-point Green function. That is, we study the
rescattering of two particles produced by the classical gluon fields,
assuming that the particles go on mass-shell both before and after the
collision. The result of this calculation appears to depend on the
ordering between the rescattering proper time $\tau_Z$ and the proper
time $\tau$ when the gluon is measured. For (i) $\tau_Z\gg 1/Q_s$ and
$\tau-\tau_Z\gg 1/Q_s$ our diagrammatic approach leads to the same
answer as that obtained by solving the Boltzmann equation. However, as
we show in Sec.~\ref{sec:free}, for (ii) $\tau-\tau_Z\gg \tau_Z\gg
1/Q_s$ the result is consistent with free-streaming gluons in the
final state, and is very different from the solution of the Boltzmann
equation. Further discussion of the physics behind the differences of
cases (i) and (ii) is presented in Sec.~\ref{sec:disc}. The resolution
of the question of whether the assumption (i) or assumption (ii) is
correct is done in the second paper \cite{KovchegovWu} of this duplex.


\section{The Schwinger-Keldysh formalism for heavy ion collisions}


\label{sec:sk}

In this Section, we shall give a detailed description of the formalism
used in our calculation. We adapt the Schwinger-Keldysh formalism
\cite{Schwinger:1960qe,Keldysh:1964ud} to describe the collision of
two particles composed of a finite number of constituents. Following
\cite{Mueller:1989st,McLerran:1993ni,McLerran:1993ka,Kovchegov:1996ty},
the two colliding nuclei are taken to consist respectively of $A_{1}$
and $A_{2}$ constituent quarks at $t=-\infty$, each valence quark
representing a nucleon. Classical gluon fields resum the parameters
$\as^2 \, A_1^{1/3}$ and $\as^2 \, A_2^{1/3}$ to all orders
\cite{Kovchegov:1997pc}: in the actual calculations below we assume
these parameters to be small, which would allow us to expand in them
perturbatively.


\subsection{The Schwinger-Keldysh formalism in perturbation theory}

\begin{figure}
\begin{center}
\includegraphics[width=0.5\textwidth]{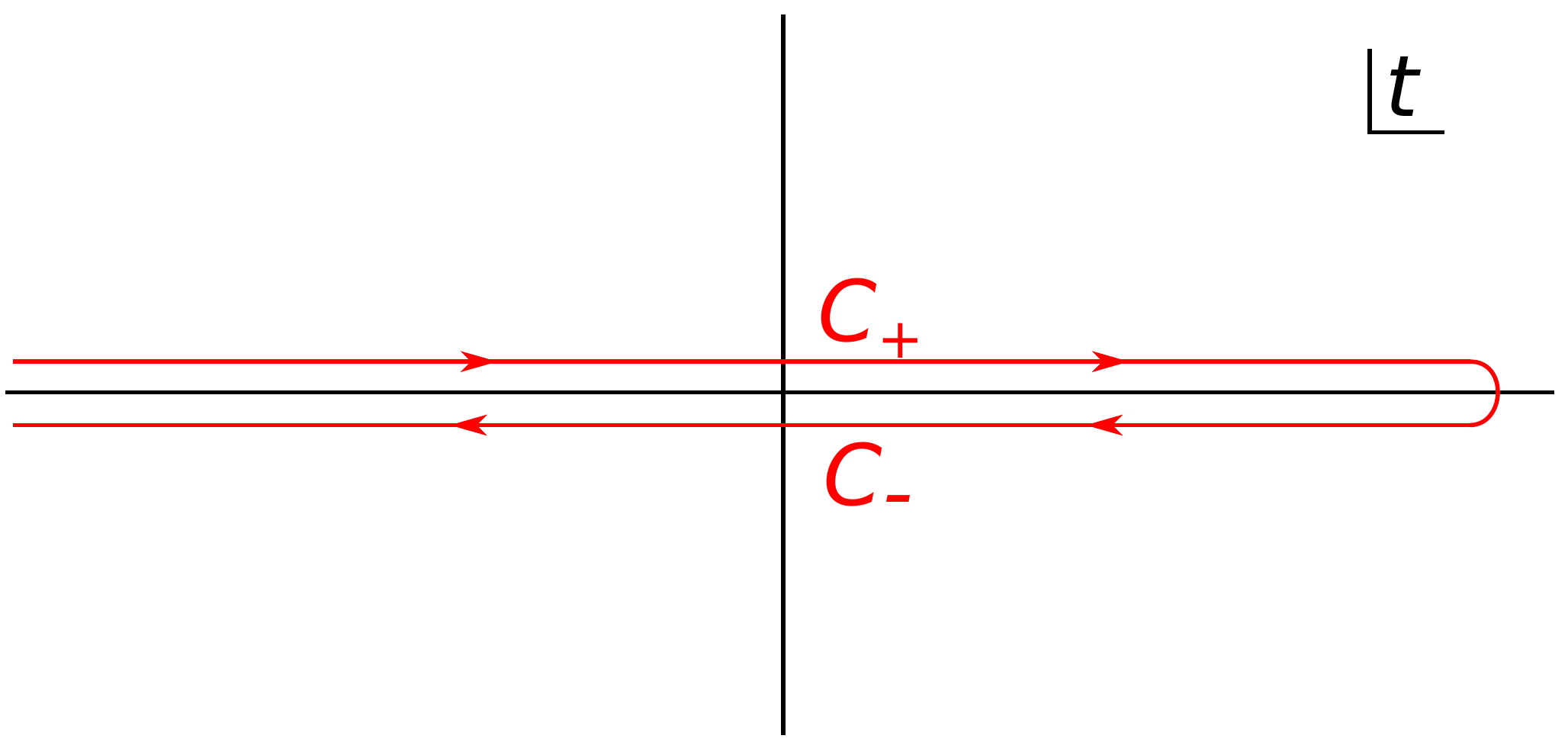}
\end{center}
\caption{The Schwinger-Keldysh contour in the complex time plane.}
\label{fig:SKcontour}
\end{figure}

We formulate our problem in terms of the density matrix $\rho$, which
can be written in terms of the wave functions of the two colliding
particles $\Psi_1$ and $\Psi_2$ before the collision
\begin{align}
\rho(t_i)=\left|\Psi_1,\Psi_2\right>\left<\Psi_1,\Psi_2\right|.
\end{align} 
In the Schr\"{o}dinger picture, the expectation value of any operator
$O$ is given by
\begin{align}
  O(t) = \Tr\left[O_s e^{-iH (t-t_i)}\rho_s (t_i) e^{i
      H(t-t_i)}\right].\label{eq:OSt}
\end{align}
In order to perform perturbative calculations, we shall use the
interaction picture by separating $H$ into a free part $H_0$ and an
interaction part $H_I$. Let us denote the operator in the interaction
picture by
\begin{align}
O_I(t)=e^{iH_0 t}O_s e^{-iH_0 t}
\end{align}
and the time evolution operator by
\begin{align}
U(t,t_i)=e^{iH_0 t}e^{-iH(t-t_i)} e^{-iH_0 t_i}=Te^{i\int d^4x \mathcal{L}_{I}}
\end{align}
where $\mathcal{L}_I$ is the interaction Lagrangian corresponding to
$H_I$. From (\ref{eq:OSt}), it is easy to show that
\begin{align}
O(t)=\Tr\left[O_I(t)U(t,t_i)\rho_I(t_i)U^\dagger(t,t_i)\right],\label{eq:Ot}
\end{align}

It is convenient to define the time ordering $T_{\mathcal{C}}$ along
the Schwinger-Keldysh contour $\mathcal{C}=\mathcal{C}_+\bigcup
\mathcal{C}_-$. As illustrated in Fig. \ref{fig:SKcontour}, the
contour $\mathcal{C}$ runs from $-\infty$ to $+\infty$ and back to
$-\infty$. On $\mathcal{C}$, the time ordering $T_{\mathcal{C}}$ can
be defined in the same way as the normal time ordering by replacing
the $\theta$ function by \cite{Niemi:1983nf}
\begin{align}
\theta_{\mathcal{C}}(t_1-t_2)=\left\{
\begin{array}{ll}
\theta(t_1-t_2)&\text{if }t_1,t_2 \in \mathcal{C}_+,\\
0&\text{if }t_1 \in \mathcal{C}_+, t_2 \in \mathcal{C}_-,\\
1&\text{if }t_2 \in \mathcal{C}_+, t_1 \in \mathcal{C}_-,\\
\theta(t_2-t_1)&\text{if }t_1,t_2 \in \mathcal{C}_-.
\end{array}
\right.
\end{align}
Accordingly, one can write
\begin{align}
  O(t) = \Tr\left[T_{\mathcal{C}} \left\{ O_I(t)
      e^{i\int_{\mathcal{C}} d^4 x \mathcal{L}_I} \rho_I(t_i) \right\}
  \right]\label{eq:OtC}
\end{align}
with $\int_{\mathcal{C}} d^4 x\equiv\int_{\mathcal{C}} d x^0\int d^3{\bf x}$. 

Given any free field $\Phi$, one can define the free propagator
\begin{align}
  G^{(0)}(x_1,x_2) & \equiv\left<0|T_{\mathcal{C}}[ \Phi(x_1)\Phi(x_2)]|0\right> \nonumber\\
  &=\theta_\mathcal{C}(t_1-t_2)[\Phi_p(x_1),\Phi_n(x_2)]+(-1)^F\theta_\mathcal{C}(t_2-t_1)[\Phi_p(x_2),\Phi_n(x_1)],\label{eq:Gfree}
\end{align}
where $\Phi$ has been decomposed into positive- and negative-
frequency parts, $\Phi_p$ and $\Phi_n$. Here $F=1$ for fermions and
$F=0$ for bosons. Using this definition one can easily generalize
Wick's theorem (see, for example, \cite{Peskin:1995ev}) to the case of
contour $\mathcal{C}$ by induction, that is,
\begin{align}
  T_{\mathcal{C}}[\Phi(x_1)\Phi(x_2)\cdots \Phi(x_m)] =
  N[& \, \Phi(x_1)\Phi(x_2)\cdots \Phi(x_m) \nonumber\\
  & + \text{all possible contractions}],\label{eq:Wick}
\end{align}
where the normal ordering operator $N[\cdots]$ puts the
negative-frequency parts to the left of all the positive-frequency
parts in the product and each contraction of two fields gives
$G^{(0)}(x_1,x_2)$. The perturbative series can be generated by using
series expansion of the exponential function in (\ref{eq:Ot})
\begin{align}
  O(t) =
  \sum\limits_{n}\frac{1}{n!}\left<\Psi_1,\Psi_2\right|T_{\mathcal{C}}
  \left\{ O_I(t) \left(i\int_{\mathcal{C}} d^4 x
      \mathcal{L}_I\right)^n \right\} \left|\Psi_1,\Psi_2\right>.
\end{align}
With Wick's theorem in (\ref{eq:Wick}) and the propagator in
(\ref{eq:Gfree}), the above equation allows one to calculate $O(t)$
perturbatively. Fields which are not contracted with other fields are
to be contracted with either $\left<\Psi_1,\Psi_2\right|$ or
$\left|\Psi_1,\Psi_2\right>$.


\subsection{QCD on the Schwinger-Keldysh contour}

With the gauge fixing term, the QCD Lagrangian in $n\cdot A=A^+=0$
light-cone gauge takes the form
\begin{align}\label{eq:L}
  \mathcal{L} = -\frac{1}{4}F^a_{\mu\nu}F^{a\mu\nu}+\bar q
  i\slashed{D} q-\frac{1}{2\xi}(n\cdot A)^2
\end{align}
with
\begin{align}
  D_\mu\equiv \partial_\mu-igA_\mu,\qquad F_{\mu\nu}=\partial_\mu
  A_\nu-\partial_\nu A_\mu-ig[A_\mu,A_\nu].
\end{align}
$\mathcal{L}$ can be separated into the free part
\begin{align}
  \mathcal{L}_0=\frac{1}{2}
  A_\mu\left[g^{\mu\nu}\partial^2-\partial^\mu\partial^\nu-\frac{n^\mu
      n^\nu}{\xi}\right]A_\nu+\bar q i\slashed{\partial} q
\end{align} and the interaction part 
\begin{align}
  \mathcal{L}_I=- gf^{abc} \partial_\mu A_\nu^a A^{b\mu}
  A^{c\nu}-\frac{g^2}{4} f^{abc} f^{ade} A_\mu^b A_\nu^c A^{d\mu}
  A^{e\nu}+g\bar q \slashed{A} q.\label{eq:LI}
\end{align}

In perturbative calculations, it is convenient to write the time
integration of $\mathcal{L}_I$ over $\mathcal{C}$ in (\ref{eq:Ot}) as
\begin{align}
  \int_{\mathcal{C}} d^4 x \, \mathcal{L}_I=\int d^4 x \,
  \mathcal{L}_I(\Phi_+) - \int d^4 x \, \mathcal{L}_I(\Phi_-)
\end{align} 
where the field $\Phi$ represents any field in $\mathcal{L}_I$ and the
subscripts $\pm$ stand for the field $\Phi$ on $\mathcal{C}_+$ and
$\mathcal{C}_-$ respectively. In the same notation as
\cite{Epelbaum:2014yja}, we shall use the retarded/advanced basis in
terms of the following fields
\begin{align}
  &\Phi_2\equiv \frac{1}{2}(\Phi_+ + \Phi_-),\qquad \Phi_1\equiv
  \Phi_+ - \Phi_- ,
\end{align}
with $\Phi_\pm$ the fields respectively on $\mathcal{C}_+$ and
$\mathcal{C}_-$ contour. That is, in order to avoid dealing with the
integration over $\mathcal{C}$, one can double the number of fields
instead. Accordingly, the propagator $G(x_1,x_2)$ can be taken as a
$2\times 2$ matrix in the space of the $1,2$ labels. In momentum
space, for a scalar field with mass $m$ we have
\begin{align}\label{eq:Gra}
G^{(0)}(k,m) =
\left(\begin{array}{cc}
0&\frac{i}{p^2-m^2-ip^0 \epsilon}\\
\frac{i}{p^2-m^2+ip^0 \epsilon}&\pi\delta(p^2-m^2)
\end{array}\right)
\equiv
\left(\begin{array}{cc}
0&G_A(p,m)\\
G_R(p,m)&G_S(p,m)
\end{array}\right) .
\end{align}
For the quark field the propagator is ($i,j$ are color indices)
\begin{align}
S_{ij}^{(0)}(k,m)=(\slashed{k}+m) \, \delta_{ij} \, G^{(0)} (k,m),\label{eq:Gq}
\end{align}
and, for the gluon field in the limit $\xi\to0$,
\begin{align}
  G^{(0)a\mu,b\nu}(k)=\left(-g^{\mu\nu}+\frac{k^\mu n^\nu+k^\nu
      n^\mu}{n\cdot k}\right)\delta^{ab} G^{(0)}(k,0).\label{eq:Gg}
\end{align}
Perturbative calculations in QCD can be carried out using the
interaction Lagrangian \cite{Jeon:2013zga}
\begin{align}\label{eq:LSK}
  \mathcal{L}_{I} \equiv & \mathcal{L}_{I} (A_+) - \mathcal{L}_{I} (A_-) = -g f^{abc}\partial_\mu\eta^a_\nu A^{b\mu} A^{c\nu}-g f^{abc}\partial_\mu A^a_\nu \eta^{b\mu} A^{c\nu}-g f^{abc}\partial_\mu A^a_\nu A^{b\mu} \eta^{c\nu}\nonumber\\
  &-\frac{g}{4} f^{abc}\partial_\mu \eta^a_\nu \eta^{b\mu} \eta^{c\nu}-g^2 f^{abc}f^{ade} \eta^b_\mu A^c_\nu A^{c\mu} A^{d\nu}-\frac{g^2}{4} f^{abc}f^{ade} \eta^b_\mu \eta^c_\nu \eta^{c\mu} A^{d\nu}\nonumber\\
  &+g\bar q_1 \slashed{A} q_2+g\bar q_2 \slashed{A} q_1+g\bar q_2
  \slashed{\eta} q_2+\frac{g}{4}\bar q_1 \slashed{\eta} q_1
\end{align}
with $A^\mu\equiv A_2^\mu$ and $\eta^\mu\equiv A_1^\mu$.

In summary, perturbative calculations of any operator can be carried
out in momentum space by the following steps:
\begin{enumerate}
\item Draw all the Feynman diagrams at a certain order in $g$ using
  the QCD vertices in (\ref{eq:LI}).

\item Assign ``1"s and ``2"s to the fields at each vertex. All the
  allowed assignments have either one or three ``1" fields at each
  vertex (see \eqref{eq:LSK}). Keep in mind that (a) the contraction
  of any two ``1" fields is always zero; and (b) the incoming states
  in the wave function of the colliding particles are only contracted
  with ``2" fields. Therefore, each external line is assigned an index
  ``2". The contraction results in spinors for external quarks and
  polarization vectors for external gluons in agreement with the
  conventional perturbative QCD.

\item Each contraction of any two fields gives the free propagator as
  one of the matrix elements of either (\ref{eq:Gq}) or
  (\ref{eq:Gg}). Here, the assignment of ``1" and ``2" gives the
  indices of the matrix element.

\item Each vertex is given by the corresponding one in the
  conventional perturbative QCD (see, say, \cite{KovchegovLevin}) with
  an overall prefactor $1/2^{n_1-1}$ with $n_1$ the number of ``1"
  fields in this vertex.

\item There is a conservation of 4-momentum at each vertex.

\item Integrate over each undetermined loop momentum.

\item Figure out the overall symmetric factor of each diagram with a
  given assignment of ``1"s and ``2"s.

\end{enumerate}

The above Feynman rules from steps 1 and 2 can be also obtained
directly by using the Lagrangian with the doubled fields in
(\ref{eq:LSK}).


\subsection{Modeling the nuclear wave function at $t=-\infty$}
\label{eq:DMinit2}

To describe heavy ion collisions we need to augment the above Feynman
rules by a specific definition of the density matrix. In this
subsection, we take the same nuclear wave function at $t=-\infty$ as
those in
Refs. \cite{Mueller:1989st,McLerran:1993ni,McLerran:1994vd,Kovchegov:1996ty}. Big
nuclei are taken to be composed of valence quarks at
$t=-\infty$. These quarks are confined in nucleons, which are
homogeneously distributed inside the nuclei with a radius $R$. We
shall study the collision of two big nuclei in the center-of-mass
frame. Partons from nucleus 1 and 2 respectively have a large ``$+$''
and ``$-$'' momenta ($v^\pm = (v^0 \pm v^3)/\sqrt{2}$), that is, these
partons are approximately moving along their respective
light-cones. The two nuclear wave functions are products of the wave
functions of nucleons, which, in turn, are products of the valence
quark wave functions.

The density matrix is
\begin{align}
  \label{eq:DMinit1}
  \rho_I(t_i)= | A_1, A_2 \rangle \langle A_1, A_2 |. 
\end{align}
We take the contribution to the density matrix coming from the ``+''
moving nucleus $A_1$ and write
\begin{align}
  \label{eq:DMA1}
  | A_1 \rangle \langle A_1 | = \prod_{i=1}^{A_1} \int \frac{d^2 p_i
    \, dp_i^+}{(2\pi)^3\, 2p_i^+} \frac{d^2 p'_i \, dp_i^{\prime
      +}}{(2\pi)^3\, 2p_i^{\prime +}} | \un{p}_i, p_i^+, k_i \rangle
  \langle \un{p}_i, p_i^+, k_i | A_1 \rangle \langle A_1 | \un{p}'_i,
  p_i^{\prime +}, l_i \rangle \langle \un{p}'_i, p_i^{\prime +}, l_i |
\end{align}
with the valence quark states $| \un{p}_i, p_i^+, k \rangle$. Here
$k_i, l_i$ are the quark color indices: summation is assumed over
repeated indices. Define the Wigner distribution of a valence quark
from nucleon $i$ in nucleus $A_1$ by
(cf. \cite{Kovchegov:2013cva,Kovchegov:2015zha})
\begin{align}\label{Wigner}
  \frac{\delta_{kl}}{N_c} \, W \left( \frac{p_i + p'_i}{2}, b_i
  \right) = \int \frac{d^2 (p_i - p'_i) \, d(p_i^+ - p_i^{\prime
      +})}{(2\pi)^3\, (p_i^+ + p^{\prime \, +}_i)} & \ e^{- i (p_i^+ -
    p_i^{\prime +}) \, b_i^- + i (\un{p}_i - \un{p}'_i) \cdot
    \un{b}_i} \notag \\ & \times \, \langle \un{p}_i, p_i^+, k | A_1
  \rangle \langle A_1 | \un{p}'_i, p_i^{\prime +}, l \rangle
\end{align}
where $p_i = (p_i^+, \un{p}_i)$ and $b_i = (b_i^-,
\un{b}_i)$. Substituting \eq{Wigner} back into \eq{eq:DMA1} we obtain
\begin{align}\label{eq:DMA2}
  & | A_1 \rangle \langle A_1 | = \prod_{i=1}^{A_1} \int \frac{d^2 P_i
    \, dP_i^+}{(2\pi)^3\, 2 P_i^+} \int d^2 b_i \, d b_i^- \, W \left(
    P_i, b_i \right) \, \frac{1}{N_c} \notag \\ & \times \, \int d^2
  (p_i - p'_i) \, d(p_i^+ - p_i^{\prime +}) \, e^{i (p_i^+ -
    p_i^{\prime +}) \, b_i^- - i (\un{p}_i - \un{p}'_i) \cdot
    \un{b}_i} \, | \un{p}_i, p_i^+, k_i \rangle \langle \un{p}'_i,
  p_i^{\prime +}, k_i |,
\end{align}
where $P_i = (p_i + p'_i)/2$. We have also approximated $p_i^+ \approx
p_i^{\prime \, +} \approx P_i^+$ since all the valence quarks in a
relativistic nucleus have approximately the same light-cone momenta.

The averaging of an operator $\hat {\cal O}$ in the state $| A_1
\rangle$ gives
\begin{align}\label{eq:DMA3}
  \langle A_1 | \hat {\cal O} | A_1 \rangle = \prod_{i=1}^{A_1} \int
  \frac{d^2 P_i \, dP_i^+}{(2\pi)^3\, 2 P_i^+} \int d^2 b_i \, d b_i^-
  \, \frac{1}{N_c} \, W \! \left( P_i, b_i \right) \, {\cal O} \left(
    \{ P_i \}, \{ b_i \} \right)
\end{align}
where
\begin{align}
  {\cal O} \left( \{ P_i \}, \{ b_i \} \right) = \prod_{i=1}^{A_1}
  \int d^2 (p_i - p'_i) \, d(p_i^+ - p_i^{\prime +}) & \, e^{i (p_i^+
    - p_i^{\prime +}) \, b_i^- - i (\un{p}_i - \un{p}'_i) \cdot
    \un{b}_i} \notag \\ & \times \, \langle \un{p}'_i, p_i^{\prime +},
  k_i | \hat {\cal O} | \un{p}_i, p_i^+, k_i \rangle
\end{align}
and the curly brackets in the argument imply dependence on all the
momenta or coordinates, e.g., $\{ P_i \} = P_1, P_2, \ldots ,
P_{A_1}$.

In the standard MV model for a large unpolarized nucleus one usually
neglects the transverse momenta $\un{P}_i$ of the valence quarks in
the nucleons and assumes that the longitudinal momentum of the nucleus
is evenly distributed among the nucleons. The corresponding
quasi-classical Wigner function in the MV model is
\cite{Kovchegov:2013cva}
\begin{align}
  \label{eq:WigMV}
  W_{cl} \left( p, b \right) = \frac{1}{A} \, \rho (b^-, \un{b}) \, 2
  (2 \pi)^3 \, \delta \left( p^+ - \frac{P^+}{A} \right) \, \delta^2
  (\un{p})
\end{align}
with the nucleon number density $\rho (b^-, \un{b})$ normalized such
that
\begin{align}
  \label{eq:density}
  \int d^2b \, d b^- \, \rho (b^-, \un{b}) = A 
\end{align}
and $P^+$ the light-cone momentum of the entire nucleus. Substituting
\eq{eq:WigMV} into \eq{eq:DMA3} we arrive at
\begin{align}
  \label{eq:DMA4}
  \langle A_1 | \hat {\cal O} | A_1 \rangle = \prod_{i=1}^{A_1}
  \sum_{k} \int d^2 b_i \, d b_i^- \, \frac{1}{A_1} \, \rho_1 (b_i^-,
  \un{b}_i) \, \frac{1}{N_c} \, {\cal O} \left( \{ b_i \} \right)
\end{align}
where $\rho_1$ is the nucleon number density in nucleus $A_1$. We have
also suppressed the momenta in the argument of $\cal O$ in
\eq{eq:DMA4}: it is understood that $\un{P}_i =0$ and $P^+_i =
P^+/A_1$ for all the nucleons (or valence quarks) in the nucleus
$A_1$.

Since for a large nucleus in the MV model $| A_1, A_2 \rangle = | A_1
\rangle \otimes | A_2 \rangle$ we conclude that the average over the
initial states of a given operator $\cal O$, which may represent a
Feynman diagram, is
\begin{align}
  \Tr \left[\rho_I (t_i) \, {\cal O} \right] = \langle A_1, A_2 |
  {\cal O} | A_1, A_2 \rangle = \prod_{i=1}^{A_1} \int d^2 b_i \, d
  b_i^- \, \frac{1}{A_1} \, \rho_1 (b_i^-, \un{b}_i) \, \frac{1}{N_c}
  \notag \\ \times \, \prod_{j=1}^{A_2} \int d^2 b'_j \, d b_j^{\prime
    \, +} \, \frac{1}{A_2} \, \rho_2 (b_j^{\prime \, +}, \un{b}'_j) \,
  \frac{1}{N_c} \, {\cal O} \left( \{ b_i \}, \{ b'_j \}
  \right) \label{eq:DMinit3}
\end{align}
where $b'_j = (b_j^{\prime \, +}, \un{b}'_j)$ are the positions of
valence quarks in the nucleus $A_2$ while $\rho_2$ is the nucleon
number density in that nucleus. 

We see that the averaging in the nuclear wave functions in the MV model
amounts only to averaging over positions and colors of the valence
quarks in the two colliding nuclei
\cite{McLerran:1993ni,McLerran:1993ka,McLerran:1994vd,Kovchegov:1996ty}. 

In the following calculations, which are leading-order in $A_1$ and
$A_2$ since they involve only one nucleon out of each nucleus, for
simplicity we will put
\begin{align}
  \label{eq:rhoAA}
  \rho_1 (b^-, \un{b}) = \frac{A}{S_\perp} \, \delta (b^-) \, \theta
  (R - b_\perp), \ \ \ \rho_2 (b^+, \un{b}) = \frac{A}{S_\perp} \,
  \delta (b^+) \, \theta (R - b_\perp).
\end{align}
We will assume that the nuclei are identical, $A_1 = A_2$, and have
the same radii. Here $R$ is the transverse radius of the nuclei and
$S_\perp = \pi R^2$ is the transverse cross-sectional area.



\section{Classical fields and quasi-particles}
\label{sec:quasi}


In this Section we calculate $G_{22}^{a\mu,b\nu}$ in the Wigner
representation
\begin{align}\label{eq:Gxp}
  G^{a\mu,b\nu}(X,p)&\equiv\int d^4x e^{ip\cdot x} G^{a\mu,b\nu}\left(X+\frac{x}{2},X-\frac{x}{2}\right)\nonumber\\
  &=\int\frac{d^4K}{(2\pi)^4} e^{-iK\cdot X}
  G^{a\mu,b\nu}\left(\frac{K}{2}+p,\frac{K}{2}-p\right)
\end{align}
at $O(A^{\frac{2}{3}} g^6)$. In thermal field theory, the free
correlation function is $G_{22}(X,p)=2\pi(n_B+1/2) \, \delta(p^2)$
with $n_B$ the Bose-Einstein distribution \cite{Niemi:1983nf}. In
systems near thermal equilibrium, one may neglect the dissipation near
the quasi-particle peak in the spectral function and take
$G_{22}(X,p)=2\pi(f+1/2) \, \delta(p^2)$ with $f$ the distribution
function in order to derive the Boltzmann equation
\cite{Chou:1984es,Calzetta:1986cq, Arnold:2002zm}. In this Section, we
shall study how the (quasi-)particle picture with $p^2=0$ emerges from
the classical fields.


\subsection{The classical field approximation at $O(A^{\frac{2}{3}}  g^6)$}

\begin{figure}[tbp]
\begin{center}
\includegraphics[width=0.6\textwidth]{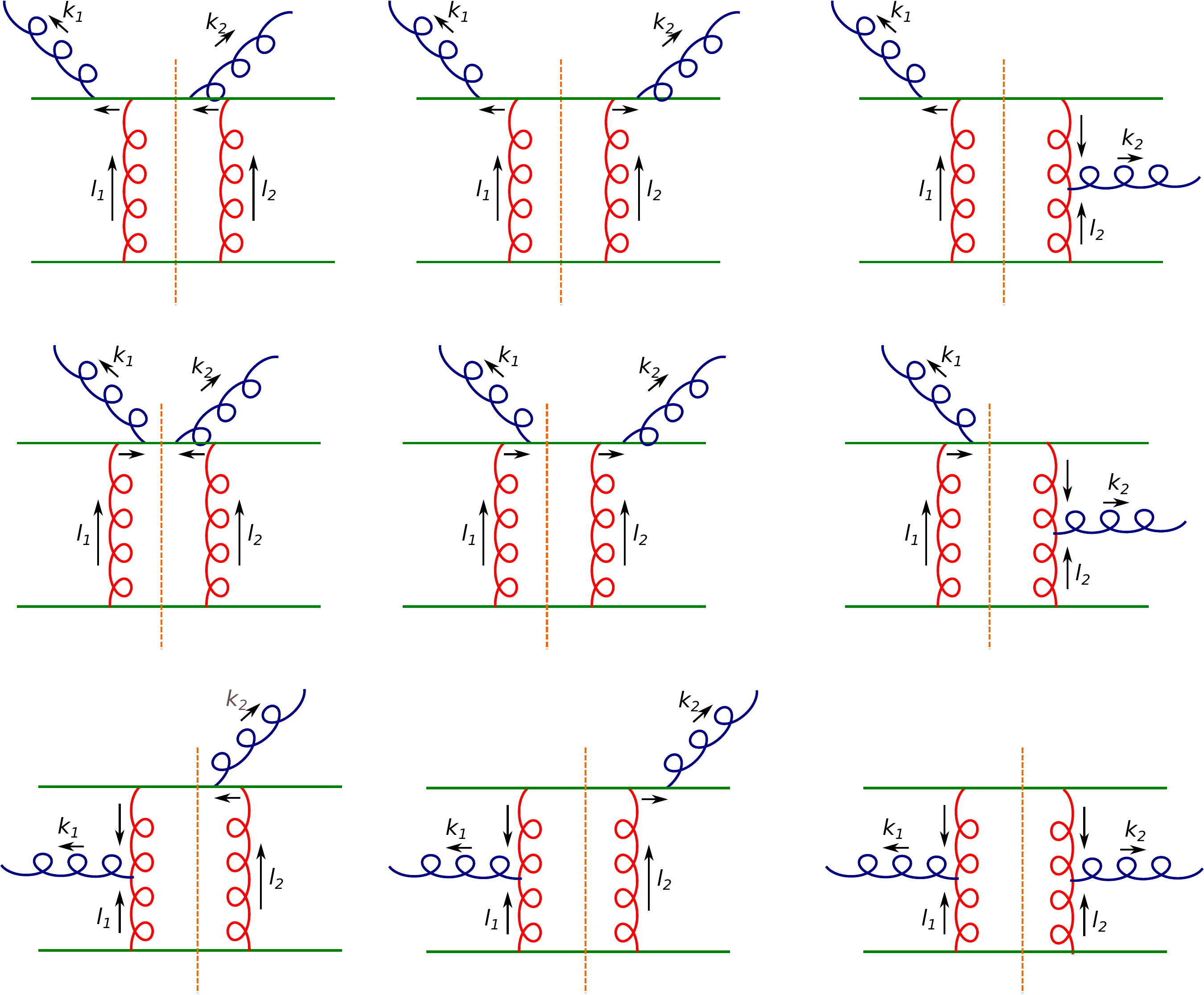}
\end{center}
\caption[*]{Diagrams at $O(A^{\frac{2}{3}}g^6)$. The $S^{(0)}_{22}$
  propagators, crossed by orange dashed lines, separate each diagram
  into two. In $A^+=0$ gauge, there are three diagrams for the
  classical field $A^{a\mu}$. Each diagram in this figure corresponds
  to that in the product of two classical fields. In each diagram the
  quark on the top has a large $P^+$ while the one at the bottom has a
  large $P^-$.}
\label{fig:g22nlo}
\end{figure}

In this Subsection, we calculate $G_{22}^{a\mu,b\nu} (k_1, k_2)$ at
$O(A^\frac{2}{3} g^6)$. The lowest-order classical gluon field in
covariant gauge was found before in
\cite{Kovchegov:1997ke,Kovchegov:2005ss}. $G_{22}^{a\mu,b\nu}$ is a
gauge-dependent quantity. We shall show that it takes a much simpler
form in $A^+=0$ gauge, which has a more transparent physical
interpretation.

We need only to evaluate the 9 diagrams\footnote{Here, we discard
  terms proportional to $\delta(x_{1,2}^\pm)$ in
  $G_{22}^{a\mu,b\nu}(x_1,x_2)$. Otherwise, there will be more
  diagrams. For example, one can not neglect the diagrams with the
  outgoing gluons attached to the quark at the bottom even in $A^+=0$
  gauge when calculating the correlation function on the light cone.}
as shown in Fig. \ref{fig:g22nlo}. In each diagram in this figure, the
quarks are put on mass shell by each $S^{(0)}_{22}$ propagator. As a
result, each diagram corresponds to that in the product of two
classical fields, in accordance with the discussion in
Appendix~\ref{sec:class}. By including all the diagrams with possible
crossing of internal gluon lines, we get (for the two identical nuclei
described in Sec.~\ref{eq:DMinit2})
\begin{align}\label{G22cl1}
  G_{22}^{a\mu,b\nu}(k_1,k_2)=\left(\frac{A}{S_\perp}\right)^2\frac{(2\pi)^2\delta(\underline{k}_1+\underline{k}_2)}{N_c^2}
  \int\frac{d^2 \underline{l}_1}{(2\pi)^2}\Tr[A_{cl}^{a\mu}(k_1,l_1)
  A_{cl}^{b\nu}(k_2,l_2)],
\end{align}
where the trace, as defined in \eqref{eq:DMinit3}, puts
$\un{l}_2=-\un{l}_1$. The classical field is
\begin{align}
  A_{cl}^{a\mu}(k,l)=&(ig)^3([T^a,T^b])(T^b)G_R(k)\frac{1}{\underline{l}^2 (\underline{l}-\underline{k})^2}\nonumber\\
  &\times\left(0, \frac{2}{k^+}
    \underline{l}\cdot(\underline{l}-\underline{k}),
    2(\underline{l}-\underline{k}) +
    \frac{(\underline{k}-\underline{l})^2\underline{k}}{k^+(k^-+i\epsilon)}\right).
\end{align}
Since we are interested in the mid-rapidity region, we only need the
pole at $k^2=0$ and we can neglect the poles at $k^\pm=0$. In this
case, one can write
\begin{align}\label{eq:ALC}
  A_{cl}^{a\mu}(k,l)=&2(ig)^3([T^a,T^b])(T^b)G_R(k)\frac{1}{\underline{l}^2 (\underline{l}-\underline{k})^2}\nonumber\\
  &\times\left(0, \frac{1}{k^+}
    \underline{l}\cdot(\underline{l}-\underline{k}),
    \frac{1}{\underline{k}^2}[\underline{k}^2 \underline{l} +
    \underline{l}^2 \underline{k} -2 \underline{k}\cdot
    \underline{l}~\underline{k}]\right)
\end{align}
such that
\begin{align}
k\cdot A_{cl}(k,l)=0.
\end{align}

By keeping only the logarithmically enhanced terms after integrating
out $\underline{l}$, that is terms with $\ln (k_{1T}/\Lambda) = \ln
(k_{2T}/\Lambda)$, we have
\begin{align}\label{Gcl}
  G_{22}^{a\mu,b\nu}(k_1,
  k_2)=&-\frac{16\pi^2\alpha_s^3\delta^{ab}}{N_c}\left(\frac{A}{S_\perp}\right)^2
  (2\pi)^2\delta(\underline{k}_1+\underline{k}_2) G_R(k_1) G_R(k_2)\nonumber\\
  &\times
  \frac{1}{(\underline{k}_{1}^2)^2}\ln\frac{\underline{k}_1^2}{\Lambda^2}\sum\limits_{\lambda=\pm}\epsilon_\lambda^\mu(k_1)
  \epsilon_{\lambda}^{*\nu}(-k_2)\equiv
\begin{array}{l}
\includegraphics[width=0.2\textwidth]{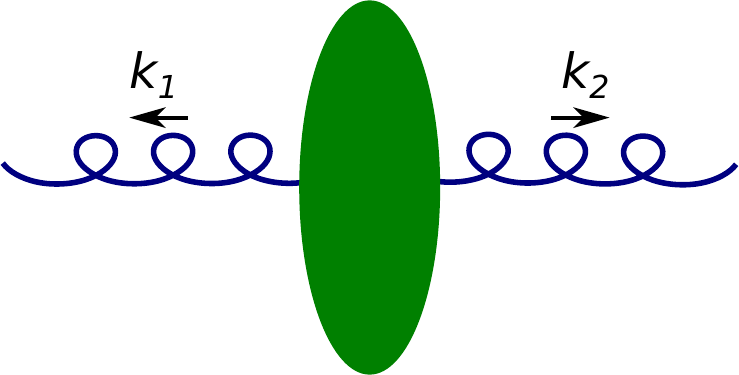}
\end{array},
\end{align}
where
\begin{align}
  \epsilon^\mu_\lambda(k)=(0,\frac{\underline{k}\cdot\underline{\epsilon}_\lambda}{k^+},\underline{\epsilon}_\lambda)\qquad\text{and}\qquad
  \sum\limits_{\lambda=\pm}\underline{\epsilon}_\lambda^i
  \underline{\epsilon}^{*j}_\lambda=\delta^{ij},
\end{align}
and $\Lambda$ is the infrared cutoff. This is much simpler than that
in covariant gauge and we have checked that it gives exactly the same
energy-momentum tensor as calculated in covariant gauge in
\cite{Kovchegov:2005ss} (see also
\cite{Lappi:2006hq,Fukushima:2007ja}).


\subsection{From classical fields to quasi-particles}

We write the retarded Green function in the following way
\begin{align}
  G_R(x) = G_{21} (x) = \theta(x^+) \, \theta(x^-) \, \int \frac{d^2 k_\perp \, d
    k^+}{(2\pi)^3 \, 2k^+} e^{-ik\cdot x}\qquad\text{with
    $k^-=\frac{\underline{k}^2}{2 k^+}$},\label{eq:GRk2}
\end{align}
where we dropped the $i \epsilon$'s in all $k^+ + i \epsilon$ and
replaced them by $\theta(x^-)$ in the prefactor. (The inverse Fourier
transform would reinstate these $i \epsilon$'s due to $\theta(x^-)$.)
$G_{22}^{a\mu,b\nu}(x_1,x_2)$ can be expressed in the following form
\begin{align}
  G_{22}^{a\mu,b\nu}(x_1, x_2)=&-\frac{16\pi^2\alpha_s^3\delta^{ab}}{N_c}\left(\frac{A}{S_\perp}\right)^2 \, \theta(x_1^+) \, \theta(x_2^+) \, \theta(x_1^-) \, \theta(x_2^-) \int \frac{d^2\underline{k}_1 dk_1^+}{(2\pi)^3 2k_1^+}  \frac{dk_2^+}{4\pi k_2^+} \nonumber \\
  &\times e^{-i k_1\cdot x_1-i k_2\cdot x_2}
  \frac{1}{(\underline{k}_{1}^2)^2}\ln\frac{\underline{k}_1^2}{\Lambda^2}\sum\limits_{\lambda=\pm}\epsilon_\lambda^\mu(k_1)
  \epsilon_{\lambda}^{*\nu}(k_2)\qquad\text{with
  }\underline{k}_2=-\underline{k}_1.
\end{align}
Inserting the above expression into the first line of (\ref{eq:Gxp})
and integrating out $\un{x}$ gives
\begin{align}
  &G_{22}^{a\mu,b\nu}(X, p)=-\frac{16\pi^2\alpha_s^3\delta^{ab}}{N_c}\left(\frac{A}{S_\perp}\right)^2 \theta(X^+) \, \theta(X^-) \int \frac{dk_1^+}{4\pi k_1^+}  \frac{dk_2^+}{4\pi k_2^+} e^{-i (k_1^+ + k_2^+) X^- -i (k_1^- + k_2^-) X^+}\nonumber\\
  &\times 
  \int\limits_{-2 X^+}^{2 X^+} d x^+ e^{i \,
    \left(p^--\frac{k_1^--k_2^-}{2}\right) x^+ } \int\limits_{-2
    X^-}^{2 X^-} d x^- e^{i \, \left(p^+ -
      \frac{k_1^+-k_2^+}{2}\right) x^-} \,
  \frac{\ln\left(\frac{\underline{p}^2}{\Lambda^2}\right)}{(\underline{p}^2)^2}
  \sum\limits_{\lambda=\pm}\epsilon_\lambda^\mu(k_1)
  \epsilon_{\lambda}^{*\nu}(k_2)
\end{align}
with $\underline{k}_1=-\underline{k}_2=\underline{p}$.

At large $X^+$ and $X^-$ one is allowed to make the following approximations
\begin{subequations}\label{late_time}
\begin{align}
  \int\limits_{-2 X^+}^{2 X^+} d x^+ e^{i \, \Delta p^- \, x^+ } \ \
  \overset{X^+ \to + \infty}{\longrightarrow} \ \
  \int\limits_{-\infty}^{\infty} d x^+ e^{i \, \Delta p^- \, x^+ } =
  2\pi\delta(\Delta p^-); \\
  \int\limits_{-2 X^-}^{2 X^-} d x^- e^{i \, \Delta p^+ \, x^- } \ \
  \overset{X^- \to + \infty}{\longrightarrow} \ \
  \int\limits_{-\infty}^{\infty} d x^- e^{i \, \Delta p^+ \, x^- } =
  2\pi\delta(\Delta p^+).
\end{align}
\end{subequations}
We get
\begin{align}
  & G_{22}^{a\mu,b\nu}(X, p)\approx-\frac{16\pi^2\alpha_s^3\delta^{ab}}{N_c}\left(\frac{A}{S_\perp}\right)^2 \theta(X^+) \theta(X^-) \int \frac{dk_1^+}{4\pi k_1^+}  \frac{dk_2^+}{4\pi k_2^+} e^{-i (k_1^+ + k_2^+) X^- -i (k_1^- + k_2^-) X^+}\nonumber\\
  & \times (2\pi)^2\delta \! \left(p^+-\frac{k_1^+-k_2^+}{2}\right)
  \delta \! \left(p^--\frac{k_1^- - k_2^-}{2}\right) \frac{1}{p_T^4}
  \ln\left(\frac{p_T^2}{\Lambda^2}\right)
  \sum\limits_{\lambda=\pm}\epsilon_\lambda^\mu(k_1)
  \epsilon_{\lambda}^{*\nu}(k_2).
  \label{eq:G22nlodd}
\end{align}
The two $\delta$-functions give us two equations, which have two
solutions
\begin{align}\label{eq:kp}
k_1^+=p^+\mp \left( \frac{p^+}{2p^-} p^2\right)^{\frac{1}{2}},\qquad k_2^+=-p^+\mp \left( \frac{p^+}{2p^-} p^2\right)^{\frac{1}{2}}.
\end{align}
Accordingly, $k_{1,2}^-=\frac{\underline{p}^2}{2 k_{1,2}^+}$ are given by
\begin{align}\label{eq:km}
k_1^-=p^-\pm \left( \frac{p^-}{2p^+} p^2\right)^{\frac{1}{2}},\qquad k_2^-=-p^-\pm \left( \frac{p^-}{2p^+} p^2\right)^{\frac{1}{2}}.
\end{align}
Taking into account the above two solutions in (\ref{eq:G22nlodd})
leads to
\begin{align}\label{eq:G22LOquas}
  &G_{22}^{a\mu,b\nu}(X,
  p)\approx\frac{16\pi^2\alpha_s^3\delta^{ab}}{N_c}\left(\frac{A}{S_\perp}\right)^2\theta(X^+)
  \theta(X^-) \frac{1}{p_T^4}\ln\left(\frac{p_T^2}{\Lambda^2}\right)
  \frac{\cos  \left(c_X \sqrt{p^2}\right)}{\sqrt{2 p^- p^+ p^2}} \\
  &\times \left(
\begin{array}{ccc}
  0&0&0\\
  0& \frac{2 p^-}{p^+} &\left[ 2 p^- + i \left(\frac{2 p^-p^2}{p^+}\right)^{\frac{1}{2}} \tan  \left(c_X \sqrt{p^2}\right)\right]\frac{\underline{p}^i}{\underline{p}^2} \\
  0&\left[ 2 p^- -i \left(\frac{2 p^-p^2}{p^+}\right)^{\frac{1}{2}} \tan  \left(c_X \sqrt{p^2}\right)\right]\frac{\underline{p}^i}{\underline{p}^2} & \delta^{ij}  \\
\end{array}
\right) \nonumber
\end{align}
with
\begin{align}
  c_X=\frac{2 \left(p^+ X^--p^- X^+\right)}{\sqrt{2 p^- p^+}}.
\end{align}
At large $X^+$, the predominant region of the above expression locates
near $p^2\simeq 0$. By neglecting terms $\propto\tan\left(c_X
  \sqrt{p^2}\right)$ we have
\begin{align}
  G_{22}^{a\mu,b\nu}(X, p)\approx\frac{16\pi^2\alpha_s^3\delta^{ab}}{N_c}&\left(\frac{A}{S_\perp}\right)^2\theta(X^+) \theta(X^-) \frac{1}{p_T^4}\ln\left(\frac{p_T^2}{\Lambda^2}\right)\nonumber\\
  & \times \, \frac{\cos \left(2 \tau\sinh(y-\eta)
      \sqrt{p^2}\right)}{\sqrt{2 p^- p^+
      p^2}}\sum\limits_{\lambda=\pm}\epsilon_\lambda^\mu(p)
  \epsilon_{\lambda}^{*\nu}(p),
\end{align}
where
\begin{align}\label{eq:rapidity}
  \eta\equiv\frac{1}{2}\ln\left(\frac{x^+}{x^-}\right),\qquad
  y\equiv\frac{1}{2}\ln\left(\frac{p^+}{p^-}\right).
\end{align}
Our result can be further simplified by taking
\begin{align}
  \lim\limits_{\tau\to \infty} \left[ \tau\frac{\cos(\tau y x)}{x}
  \right] = 2 \pi \delta(y)\delta(x^2).\label{eq:dd}
\end{align}
The above equation holds because the support of the left-hand side is
limited to $x=0$ as $\tau\gg 1$ and
\begin{align}
  \int_0^\infty dx^2 \frac{\cos(\tau y x)}{x}= \int_0^\infty dx
  \left(e^{i xy}+e^{-i
      xy}\right)=\frac{1}{\tau}\left(\frac{i}{y+i\epsilon}-\frac{i}{y-i\epsilon}\right)=
  \frac{2\pi}{\tau}\delta(y).
\end{align}
As a result, we have
\begin{align}\label{eq:G22cl}
  G_{22}^{a\mu,b\nu}(X, p)\to 2\pi\delta(p^2) \delta^{ab}
  \sum\limits_{\lambda=\pm}\epsilon_\lambda^\mu(p)
  \epsilon_{\lambda}^{*\nu}(p) f^{cl}(X,p),
\end{align}
where
\begin{align}
  f^{cl}(X,p)=\frac{1}{\tau} \theta(X^+) \theta(X^-) \delta(y-\eta)
  f_\perp^{cl}(\underline{p})
  \label{eq:fcl}
\end{align}
 and
\begin{align}
  f_\perp^{cl}(\underline{p})\equiv\frac{8\pi^2\alpha_s^3}{N_c}\left(\frac{A}{S_\perp}\right)^2
  \frac{1}{p_T^5}\ln\left(\frac{p_T^2}{\Lambda^2}\right).
\end{align}

Our result in (\ref{eq:G22cl}), while obtained in the classical field
approximation, has a physical interpretation in terms of particles. We
have taken the longitudinal size of the two nuclei to be zero in
(\ref{eq:rhoAA}). As a result, they collide at $t=0=z$. After that,
each produced gluon travels at the speed of light. Along the
$z$-direction, its location $X^3=v_z X^0$ with $v_z=p_z/p^0$. This is
what leads to the $\delta$-function at $\eta=y$.

From (\ref{eq:fcl}), one can easily see that the longitudinal pressure
is zero at mid-rapidity due to $\delta(y-\eta)$. This is what has been
observed in \cite{Kovchegov:2005ss}. Numerical simulations have shown
that including all the other classical diagrams will not change the
fact that the longitudinal pressure approaches zero much faster than
the transverse pressure at late times
\cite{Krasnitz:1998ns,Krasnitz:1999wc,Krasnitz:2003nv,Lappi:2003bi,Berges:2013fga}.


\section{Rescattering and the Boltzmann equation}
\label{sec:resc}

\begin{figure}[htb]
\begin{center}
\includegraphics[width=0.8\textwidth]{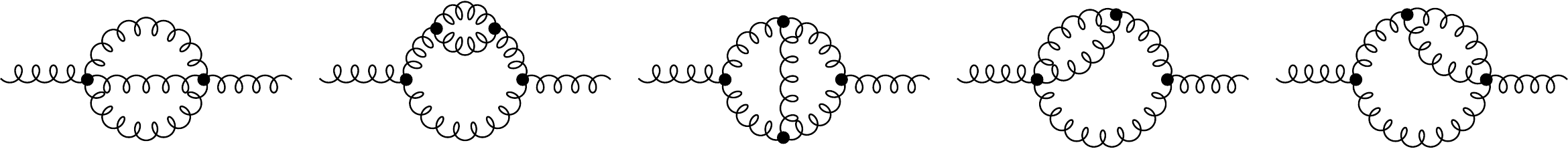}
\end{center}
\caption[*]{Two-loop diagrams for the gluon two-point function. Here,
  we only include the two-particle irreducible (2PI) gluon
  self-energies in each diagram.}
\label{fig:g22}
\end{figure}

In this Section we will use $G_{22}^{a\mu,b\nu}(X,p)$ of
$O(A^{\frac{2}{3}} g^6)$ in (\ref{eq:G22cl}) to evaluate a subset of
diagrams of $O(g^{16} A^{\frac{4}{3}})$. This subset of diagrams can
be obtained by assigning ``1"'s and ``2"'s to each diagram in
\fig{fig:g22} and replacing two of its 2-2 propagators with the
classical one in (\ref{eq:G22cl}). That is, the two of the 2-2
propagators are replaced by the 9 diagrams in \fig{fig:g22nlo} with
all the possible crossings of their internal gluon lines. We shall
show that under a certain approximation these diagrams give a result
identical to that obtained by solving the Boltzmann equation via
perturbative expansion in the collision term. In this sense they give
the contribution to $G_{22}^{a\mu,b\nu}(X,p)$ from rescattering
between the produced gluons. However, under a different approximation
these diagrams do not reduce to a solution of Boltzmann equation.

\subsection{$G_{22}^{a\mu,b\nu}(X,p)$ from rescattering}
\label{sec:Gresc}

\begin{figure}
\begin{center}
\includegraphics[width=0.8\textwidth]{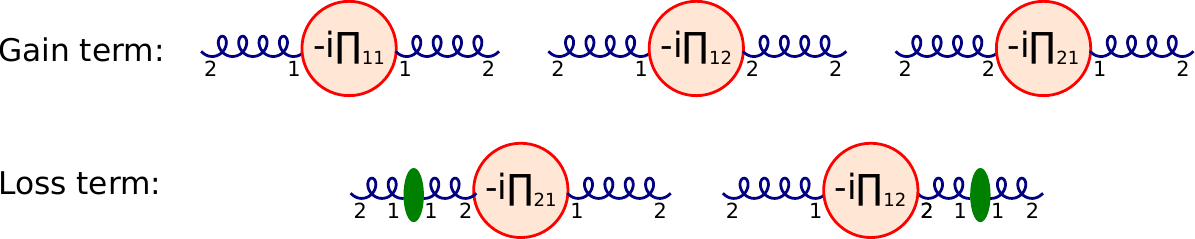}
\end{center}
\caption{Grouping the diagrams for rescattering. Here, the green ovals
  represent the diagrams at $O(A^{\frac{2}{3}} g^6)$ while the circles
  denote 2PI self-energies.}
\label{fig:g22lossgain}
\end{figure}

For the simplicity of the color and Lorentz indices, we shall
calculate
\begin{align}
  \bar G_{22}(X,p) \equiv \frac{\sum_\lambda
    \epsilon_{\lambda\nu}(p)\epsilon^*_{\lambda\mu}(p)}{2(N_c^2-1)} \,
  G_{22}^{a\mu,a\nu}(X,p).\label{eq:Gbar}
\end{align}
As illustrated in Fig.~\ref{fig:g22lossgain}, we can group the subset
of diagrams into a gain term and a loss term in the kinetic theory
notation, with the circles denoting 2PI self-energies $\Pi_{ij}$. In
the gain term two classical $G_{22}^{a\mu,b\mu}$ are used in the
calculation of the self-energies, $\Pi$'s, while in the loss term only
one classical $G_{22}^{a\mu,b\mu}$ is used in $\Pi$'s and the other
classical correlator is placed on one of the external gluon
propagators, as shown by the green ovals in \fig{fig:g22lossgain}.

In terms of the averaged self-energies
\begin{align}
  \bar{\Pi}\equiv\frac{\sum_\lambda
    \epsilon_{\lambda\nu}(p)\epsilon^*_{\lambda\mu}(p)
    \Pi^{a\mu,b\nu}(X,p)}{2 (N_c^2-1)},
\end{align}
the gain term takes the form
\begin{align}
  \bar G_{22}^{gain}(X,p)=&-i\int d^4x \, e^{ip\cdot x}\int d^4 z \, e^{-ip'\cdot z}\int\frac{d^4Z \, d^4p'}{(2\pi)^4}\nonumber\\
  &\times\left[ G^{(0)}_{21}\left(X-Z+\frac{x-z}{2}\right)\bar\Pi_{11}\left(Z,p' \right)G^{(0)}_{12}\left(Z-X+\frac{x-z}{2}\right)\right.\nonumber\\
  &\left.+ G^{(0)}_{21}\left(X-Z+\frac{x-z}{2}\right)\bar\Pi_{12}\left(Z,p' \right)G^{(0)}_{22}\left(Z-X+\frac{x-z}{2}\right)\right.\nonumber\\
  &\left.+
    G^{(0)}_{22}\left(X-Z+\frac{x-z}{2}\right)\bar\Pi_{21}\left(Z,p'
    \right)G^{(0)}_{12}\left(Z-X+\frac{x-z}{2}\right)\right]. \label{eq:G22formal}
\end{align}

To evaluate this expression we write the retarded Green function
$G^{(0)}_{21}(x)$ in the form of (\ref{eq:GRk2}), while the advanced
and cut Green functions are
\begin{align}
  G^{(0)}_{12}(x)= G_A(x)=-\theta(-x^+) \, \theta(-x^-) \int \frac{d^2
    k_\perp \, d k^+}{(2\pi)^3 \, 2k^+} \, e^{-ik\cdot
    x}\qquad\text{with $k^-=\frac{\underline{k}^2}{2
      k^+}$}\label{eq:GAk2}
\end{align}
and
\begin{align}
  G^{(0)}_{22}(x)= \int \frac{d^4 k}{(2 \pi)^4} \, e^{-ik\cdot x} \,
  \pi \, \delta (k^2) = \frac{1}{2} \int \frac{d^2 k_\perp \, d
    k^+}{(2\pi)^3 \, 2 |k^+|} \, e^{-ik\cdot x} .\label{eq:Gcut}
\end{align}
Integrating over $z$, $p'$, $\un{x}$ we arrive at
\begin{align}
  \label{eq:G22long}
  & \bar G_{22}^{gain}(X,p)=i \int \frac{d^2 k_\perp \, d
    k^+}{(2\pi)^3 \, 2k^+} \frac{4 \, d k^{\prime \, +}}{4 \pi
    k^{\prime \, +}} \, e^{-i (k-k') \cdot (X-Z)} \, d^4 Z \\ & \times
  \, \left\{ \theta (X^+ - Z^+) \theta (X^- - Z^-) \int\limits_{-2
      (X^+ - Z^+)}^{2 (X^+ - Z^+)} d y^+ e^{i \, \left(p^--\frac{k^-+
          k^{\prime \, -}}{2}\right) y^+ } \right.  \notag \\ &
  \hspace*{3cm} \times \, \int\limits_{-2 (X^- - Z^-)}^{2 (X^- - Z^-)}
  d y^- e^{i \, \left(p^+ - \frac{k^+ + k^{\prime \, +}}{2}\right)
    y^-} \, \bar\Pi_{11} (Z,P) \notag \\ & - \frac{\mbox{Sign}
    (k^{\prime \, +})}{2} \, \int\limits_{-2 (X^+ - Z^+)}^{\infty} d
  y^+ e^{i \, \left(p^--\frac{k^-+ k^{\prime \, -}}{2}\right) y^+ }
  \int\limits_{-2 (X^- - Z^-)}^{\infty} d y^- e^{i \, \left(p^+ -
      \frac{k^+ + k^{\prime \, +}}{2}\right) y^-} \, \bar\Pi_{12}
  (Z,P) \notag \\ & \left. + \frac{\mbox{Sign} (k^{+})}{2} \,
    \int\limits_{-\infty}^{2 (X^+ - Z^+)} d y^+ e^{i \,
      \left(p^--\frac{k^-+ k^{\prime \, -}}{2}\right) y^+ }
    \int\limits_{-\infty}^{2 (X^- - Z^-)} d y^- e^{i \, \left(p^+ -
        \frac{k^+ + k^{\prime \, +}}{2}\right) y^-} \, \bar\Pi_{21}
    (Z,P) \right\} \notag
\end{align}
where we have defined $y^\pm = x^\pm - z^\pm$. Here $\un{k}' = 2
\un{p} - \un{k}$.

To reproduce kinetic theory one has to assume that gluons go on mass
shell between interactions. This means the time between rescatterings
is long enough for the gluons to go on mass shell. Therefore, we need
to assume that $X^+ - Z^+$ and $X^- - Z^-$ are very large in
\eq{eq:G22long}. This approximation is different from simply assuming
that $X^+$ and $X^-$ are large, as was done in Eqs.~\eqref{late_time},
since the integrals over $Z^+$ and $Z^-$ in \eq{eq:G22long} are not
restricted to the regions far away from $X^+$ and $X^-$
respectively. Thus we simply assume that the large-$X^+ - Z^+$ and
$X^- - Z^-$ region dominates in the integral. This assumption is
needed to obtain kinetic theory from our formalism, but cannot be
easily justified otherwise for the collision at hand.

When assuming that a dimensionful quantity is large one has to compare
it to another dimensionful quantity. Unfortunately this is hard in our
case, since almost everything else is integrated out. We simply state
that $X^+ - Z^+$ and $X^- - Z^-$ are the largest distance scales in
the problem, with the possible exception of $Z^+$ and $Z^-$ which may
be comparable. Note that in deriving the classical correlator
\eqref{eq:G22cl} we have assumed that $X^+$ and $X^-$ are large (see
\eqref{late_time}): in the problem at hand, $X^+$ and $X^-$ from
\eq{eq:fcl} become $Z^+$ and $Z^-$ since we will be using the
classical correlators to calculate $\bar\Pi_{ij}$. Therefore, our
$Z^+$ and $Z^-$ have already been assumed to be very large.

Finally, a question remains whether to send $X^+ - Z^+$ and $X^- -
Z^-$ to $+\infty$ or to $-\infty$ when taking them large: from the
curly brackets in \eq{eq:G22long} we see that only the $X^+ - Z^+ \to
+\infty $ and $X^- - Z^- \to + \infty$ limits give a non-zero
result. Applying those limits to \eq{eq:G22long} with the help of
Eqs.~\eqref{late_time} and integrating out $k^+$, $k^{\prime \, +}$
and $\un{Z}$ afterwards while assuming that $\bar\Pi_{ij}(Z^+, Z^-,
\un{Z},P) = \bar\Pi_{ij}(Z^+, Z^-,P)$ due to the slowly changing
transverse profile of the large nucleus yields
\begin{align}\label{eq:G22shorter}
  \bar G_{22}^{gain}(X,p)=i\int\limits_0^{X^+} dZ^+
  \int\limits_0^{X^-} dZ^- \,
  \frac{\cos(c_{X-Z}\sqrt{p^2})}{\sqrt{2p^+p^-
      p^2}}\left[\bar\Pi_{11}+\frac{\mbox{Sign}
      (p^+)}{2}(\bar\Pi_{21}-\bar\Pi_{12})\right],
\end{align}
where
\begin{align}
  c_{X-Z}=2 \tau_{X-Z}\sinh(y-\eta_{X-Z})
\end{align}
with
\begin{align}
  \tau_{X-Z}\equiv\sqrt{2(X^+-Z^+)(X^--Z^-)},\qquad \eta_{X-Z}\equiv
  \frac{1}{2}\ln\left(\frac{X^+-Z^+}{X^--Z^-}\right).
\end{align}
In arriving at \eq{eq:G22shorter} we put $p^2=0$ in the argument of
the Sign-function: this approximation will be justified shortly. Lower
limits of the $Z^+$ and $Z^-$ integrals were set to zero in
\eq{eq:G22shorter} due to the classical correlator \eqref{eq:fcl}
which we will use to calculate $\bar\Pi_{ij}$: the correlator ensures
that no gluons are produced before the heavy ion collision at $(t,z) =
(0,0)$.

It is important to point out that, even though we assumed that $X^+ -
Z^+$ and $X^- - Z^-$ are very large, we have set the upper limits of
the $Z^+$ and $Z^-$ integrations in \eq{eq:G22shorter} to $X^+$ and
$X^-$ respectively. This is related to the fact that our calculation
requires that $X^+ - Z^+$ and $X^- - Z^-$ are large, but does not tell
us whether they need to be larger than $Z^+$ and $Z^-$. For instance,
large $X^+ - Z^+$ may imply either of the following situations (ditto
for $X^- - Z^-$):
\begin{itemize}
\item[(i)] $X^+ - Z^+ \gg 1/Q_s$, $Z^+ \gg 1/Q_s$; or
\item[(ii)] $X^+ - Z^+ \gg Z^+ \gg 1/Q_s$.
\end{itemize}
As we will see below, the two limits give different results. As we
mentioned in the Introduction, we will answer the question of whether
regime (i) or (ii) is correct by a more detailed calculation in our
next paper \cite{KovchegovWu}.

By assuming that $\tau_{X-Z}$ is sufficiently large once again and
using (\ref{eq:dd}), we obtain
\begin{align}
  \bar G_{22}^{gain}(X,p)=&\frac{i\pi}{p_\perp} \, \delta(p^2) \,
  \int\limits_0^{X^+} dZ^+ \int\limits_0^{X^-} dZ^- \, \delta(y-\eta_{X-Z}) \, \frac{1}{\tau_{X-Z}} \nonumber\\
  &\times\left[\bar\Pi_{11}(Z,p)+\frac{\mbox{Sign} (p^0)}{2}
    \left(\bar\Pi_{21}(Z,p)-\bar\Pi_{12}(Z,p)\right)\right].
\label{eq:G22gainPi}
\end{align}

Now let us turn our attention to the loss term. We will make similar
approximations while evaluating the loss term in
\fig{fig:g22lossgain}. The exact starting form of the loss term is as
follows
\begin{align}
  \bar G^{loss}_{22}(X,p)= & - i\int d^4x e^{ip\cdot x} \int d^4 z_1  d^4 z_2\nonumber\\
  & \times\left[ G^{(0)}_{21}\left(X+\frac{x}{2}-z_1\right) \, \bar\Pi_{12}(z_1,z_2) \, \bar G^{cl}_{22}\left(z_2,X-\frac{x}{2}\right) \right. \nonumber \\
  &\left.+ \bar G^{cl}_{22}\left(X+\frac{x}{2},z_1\right)
    \bar\Pi_{21}(z_1,z_2) \, G^{(0)}_{12}\left(z_2 - X +
      \frac{x}{2}\right)\right],\label{eq:G22loss}
\end{align}
where $\bar G^{cl}_{22}$ is obtained by substituting the classical
correlator $G_{22}^{a\mu,b\nu}$ from (\ref{eq:G22cl}) into
(\ref{eq:Gbar}). Similar to the above we define $Z = (z_1 + z_2)/2$,
$z = z_1 - z_2$ and write
\begin{align}
  \bar G^{loss}_{22}(X,p)= & - i\int d^4x e^{ip\cdot x} \int d^4 Z d^4
  z \frac{d^4 p'}{(2 \pi)^4} \, e^{-i p'
    \cdot z} \, \nonumber\\
  & \times\left[ G^{(0)}_{21}\left(X-Z+\frac{x-z}{2} \right) \,
    \bar\Pi_{12}(Z, p') \, \bar G^{cl}_{22}\left(Z - \frac{z}{2} , X
      -\frac{x}{2}\right) \right. \nonumber \\ &\left. + \, \bar
    G^{cl}_{22}\left(X+\frac{x}{2} , Z + \frac{z}{2} \right)
    \bar\Pi_{21}(Z, p') \, G^{(0)}_{12}\left(Z-X
      +\frac{x-z}{2}\right)\right] \label{eq:G22loss1}
\end{align}
with
\begin{align}
  \label{eq:Pi_tr}
  \bar\Pi_{ij}(z_1,z_2) = \bar\Pi_{ij} \left(Z + \frac{z}{2} , Z -
    \frac{z}{2} \right) = \int \frac{d^4 p'}{(2 \pi)^4} \, e^{-i p'
    \cdot z} \, \bar\Pi_{ij} (Z, p').
\end{align}

Integrating over $z$, $p'$ and $\un{x}$ yields
\begin{align}  \label{eq:G22loss2}
  & \bar G^{loss}_{22}(X,p)= i \int \frac{d^2 k_\perp \, d
    k^+}{(2\pi)^3 \, 2k^+} \frac{4 \, d k^{\prime \, +}}{4 \pi
    k^{\prime \, +}} \, e^{i (k-k') \cdot (X-Z)} \, d^4 Z \, f^{cl}
  \left( \frac{X+Z}{2}, k'\right) \\ & \times \, \left\{ \mbox{Sign}
    (k^{\prime \, +}) \, \int\limits_{-\infty}^{2 (X^+ - Z^+)} d y^+
    e^{i \, \left(p^--\frac{k^-+ k^{\prime \, -}}{2}\right) y^+ }
    \int\limits_{-\infty}^{2 (X^- - Z^-)} d y^- e^{i \, \left(p^+ -
        \frac{k^+ + k^{\prime \, +}}{2}\right) y^-} \, \bar\Pi_{21}
    (Z,P) \right. \notag \\ & \left. - \mbox{Sign} (k^{+}) \,
    \int\limits_{-2 (X^+ - Z^+)}^{\infty} d y^+ e^{i \,
      \left(p^--\frac{k^-+ k^{\prime \, -}}{2}\right) y^+ }
    \int\limits_{-2 (X^- - Z^-)}^{\infty} d y^- e^{i \, \left(p^+ -
        \frac{k^+ + k^{\prime \, +}}{2}\right) y^-} \, \bar\Pi_{12}
    (Z,P) \right\} \notag
\end{align}
where again $y^\pm = x^\pm - z^\pm$ along with $\un{k}' = 2 \un{p} -
\un{k}$. We have also assumed that $x^\pm$ and $z^\pm$ are much
smaller than $X^\pm$ and $Z^\pm$ and neglected $x^\pm$ and $z^\pm$ in
the argument of $f^{cl}$.

Assuming that $X^+ - Z^+ \to +\infty $ and $X^- - Z^- \to + \infty$
and $\bar\Pi_{ij}(Z^+, Z^-, \un{Z},P) = \bar\Pi_{ij}(Z^+, Z^-,P)$ we
integrate over $y^+$, $y^-$, $\un{Z}$, $k^+$ and $k^{\prime \, +}$
obtaining
\begin{align} 
  \label{eq:G22loss3}
  \bar G^{loss}_{22}(X,p)= i \int\limits_0^{X^+} dZ^+
  \int\limits_0^{X^-} dZ^- \,
  \frac{\cos(c_{X-Z}\sqrt{p^2})}{\sqrt{2p^+p^- p^2}} \, \mbox{Sign}
  (p^+) \, (\bar\Pi_{21}-\bar\Pi_{12}) \, f^{cl} \left( \frac{X+Z}{2},
    p \right)
\end{align}
where again we have put $p^2 =0$ in the argument of the Sign-function
along with the argument of $f^{cl}$.

Finally, invoking the late-time argument again we apply \eq{eq:dd} to
\eq{eq:G22loss3}. This gives
\begin{align}
  \bar G_{22}^{loss}(X,p) = & \, \frac{i\pi}{p_\perp} \, \delta(p^2) \int_0^{X^+} dZ^+ \int_0^{X^-} dZ^- \, \delta(y-\eta_{X-Z}) \frac{1}{\tau_{X-Z}} \notag \\
  & \times \mbox{Sign} (p^0) \,
  \left[\bar\Pi_{21}(Z,p)-\bar\Pi_{12}(Z,p)\right] f^{cl}\left(
    \frac{X+Z}{2}, p \right). \label{eq:G22lossPi}
\end{align}


\subsection{Gluon self-energies}

\begin{figure}
\begin{center}
\includegraphics[width=0.7\textwidth]{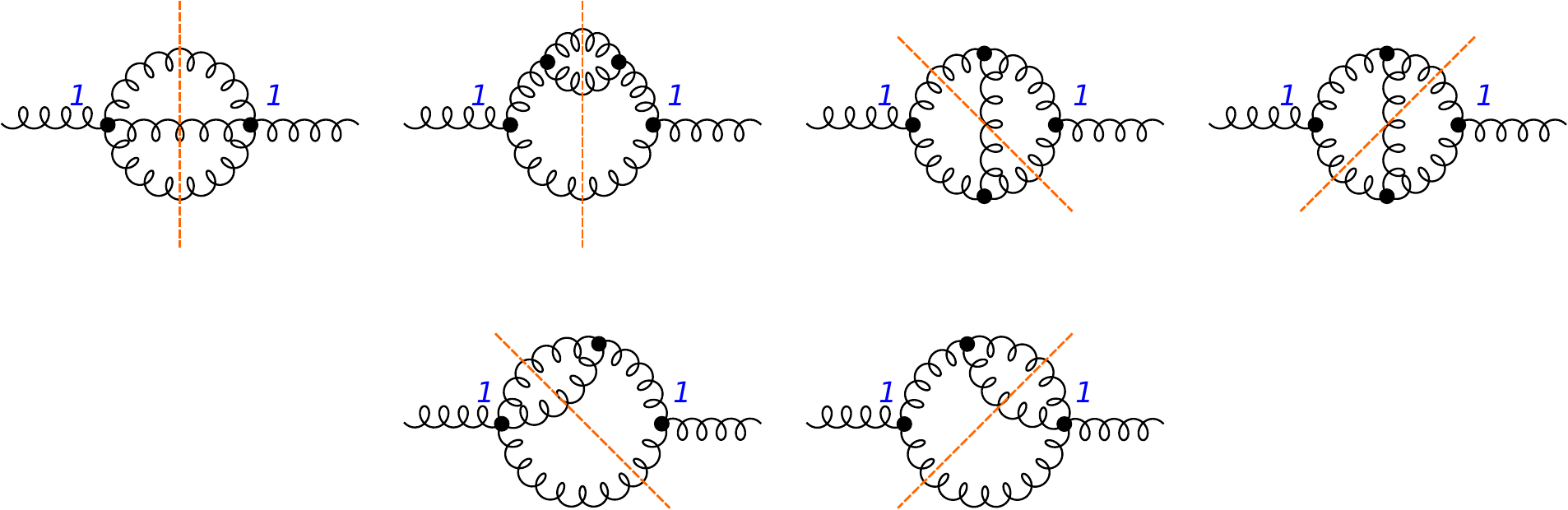}
\end{center}
\caption[*]{Diagrams for $\Pi_{11}^{a\mu,b\nu}$. In each diagram the
  dashed orange line cuts through all the $2-2$ propagators,
  separating the diagram into a product of an amplitude and a complex
  conjugate amplitude.}
\label{fig:Pi11}
\end{figure}

We first evaluate the gluon self energy $\Pi_{11}^{a\mu,b\nu}$ from
the diagrams in Fig. \ref{fig:Pi11}. Without loss of generality (for
the late-time approximation at hand), we assume that the gluon
propagators in these diagrams take the following form
\begin{align}
  &G_{22}^{a\mu,b\nu}(X,p)=2\pi \delta(p^2) \delta^{ab} \sum_\lambda \epsilon_\lambda^\mu(p) \epsilon_{\lambda}^{*\nu}(p) \, g_{22}(X,p),\nonumber\\
  &G_{21}^{a\mu,b\nu}(X,p)=G_R^{a\mu,b\nu}(p),\qquad
  G_{12}^{a\mu,b\nu}(X,p)=G_A^{a\mu,b\nu}(p),\label{eq:Ganzatz}
\end{align}
with $g_{22}(X,p)=1/2$ for the free one and $g_{22}(X,p)=f^{cl}(X,p)$
for the classical one. For our problem, we need only to include
vertices with only one ``1" field. In each diagram there are three
$2-2$ propagators according to the counting rule in (\ref{eq:pc}).

In our calculation, we choose to label by $p_1,p_2,p_3$ the momenta of
the three $2-2$ propagators in each diagram. These will be our
integration variables. They satisfy $p=p_1+p_2+p_3$. Each propagator
has a positive- and negative-frequency part. We shall take the
external momentum $p^+$ to be positive, and, in view of the above
calculation of the gain and loss term, on mass shell, $p^2 =0$. In
each diagram, while evaluating loop integrals, there should be only
two lines out of $p_1,p_2,p_3$-carrying 2-2 lines with the
positive-frequency parts of the propagators, while the remaining third
line would come in with the negative-frequency part. It is clear that
the diagrams in \fig{fig:Pi11} reduce to the $gg \to gg$ scattering
amplitude squared.  Except for the first diagram in
Fig. \ref{fig:Pi11}, different choices of positive- and
negative-frequency parts for the 2-2 lines give us the products of
$s-$, $t-$ and $u-$ channel amplitudes and their conjugates. Since
$p_1, p_2$ and $p_3$ are dummy variables to be integrated out, we
redefine $p_1$ as the negative-frequency momentum and replace $p_1 \to
- p_1$ such that the new $p_1$ would have a positive frequency. Then,
by collecting all terms obtained in this way, we arrive at the
following result
\begin{align}
  -i\bar{\Pi}_{11}&=-\frac{1}{2}\int_{p_1,p_2,p_3} (2\pi)^4
  \delta(p+p_1-p_2-p_3)
 \nonumber\\
 &\times 
\, \overline{|M|^2} \,
  g_{22}(X,p_1) \, g_{22}(X,p_2) \, g_{22}(X,p_3),
  \label{eq:Pi11}
\end{align}
where
\begin{align}
  \overline{|M|^2}= 8 N_c^2 g_s^4 \left(3-\frac{t u}{s^2}-\frac{s
      u}{t^2}-\frac{s t}{u^2}\right),\label{eq:Msgggg}
\end{align}
and for brevity we have denoted
\begin{align}
\int_p\equiv\int\frac{d^3 p}{(2\pi)^3 2 \omega_{p}},
\end{align}
with $\omega_{p} = |{\vec p}|$ and the Mandelstam variables defined by
\begin{align}
  s=(p+p_1)^2,\qquad t=(p-p_2)^2,\qquad u=(p-p_3)^2.
\end{align}

\begin{figure}[htb]
\begin{center}
\includegraphics[width=0.7\textwidth]{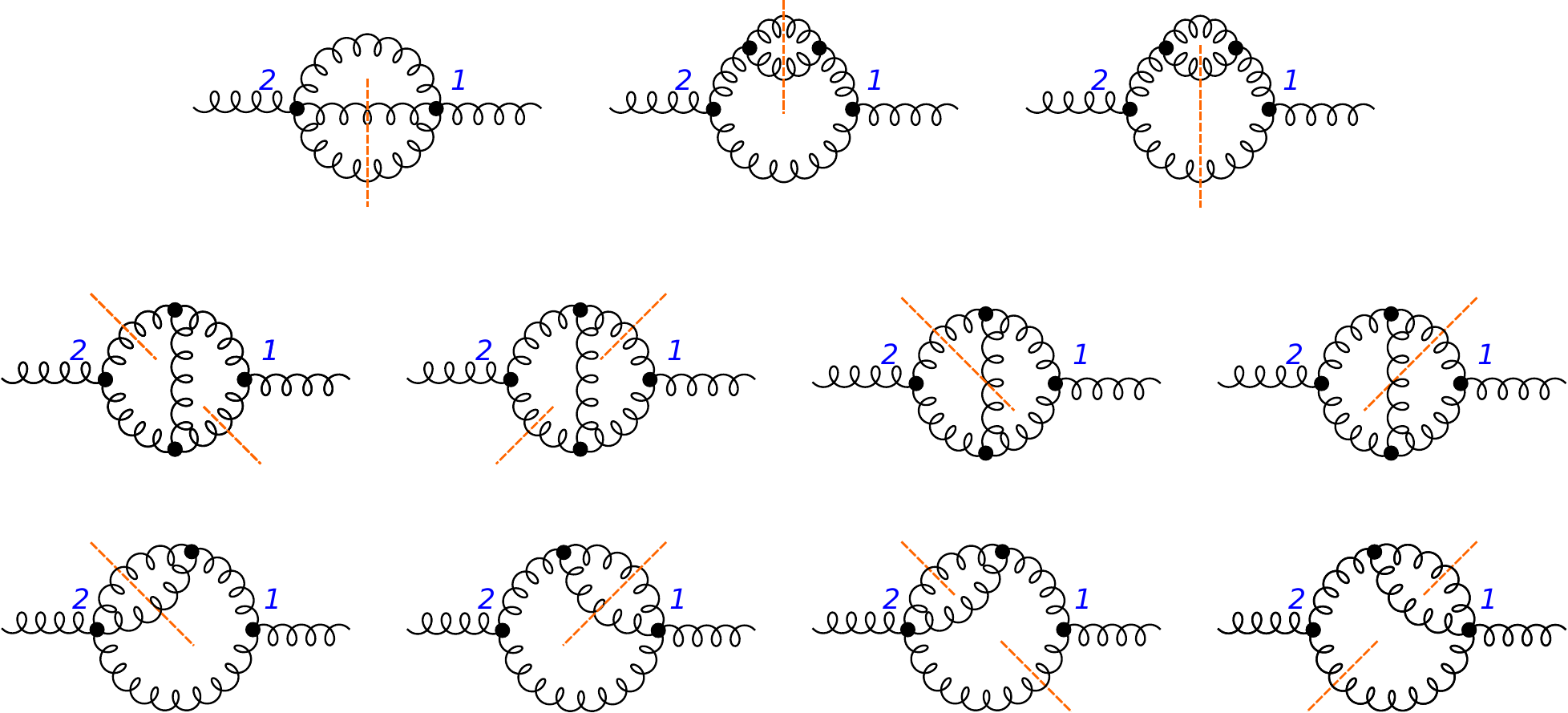}
\end{center}
\caption[*]{Diagrams for $-i \Pi_{21}^{a\mu,b\nu}$. In each diagram the
  dashed orange line cuts through two $2-2$ propagators.}
\label{fig:Pi21}
\end{figure}

Next, let us calculate $-i [\bar\Pi_{21}-\bar\Pi_{12}]$, which are
given by 2 times the real parts of the diagrams in
Fig. \ref{fig:Pi21}. As indicated by the dashed lines in this figure,
there are only two $2-2$ propagators in each diagram. Compared to each
corresponding diagram for $\bar\Pi_{11}$ in Fig. \ref{fig:Pi11}, the
diagrams in \fig{fig:Pi21} have a retarded (or advanced) propagator
instead of the third $2-2$ propagator. Then, by subtracting out
$\bar\Pi_{12}$ from $\bar\Pi_{21}$ one converts the retarded
(advanced) propagator into a on-mass shell $\delta$-function with
different signs for its positive- and negative-frequency parts. After
this, using the same trick as that for $\Pi_{11}$ with the positive
and negative energy parts of the propagators, we get
\begin{align}
  -i&[\bar\Pi_{21}(X,p)-\bar\Pi_{12}(X,p)]
  =\frac{1}{2}\int_{p_1,p_2,p_3} (2\pi)^4\delta(p+p_1-p_2-p_3) \, \overline{|M|^2}  \nonumber\\
  & \times \, \left[g_{22}(X,p_1) \, g_{22}(X,p_2) + g_{22}(X,p_1) \,
    g_{22}(X,p_3) - g_{22}(X,p_2) \, g_{22}(X,p_3) \right] .
  \label{eq:Pi21}
\end{align}

\subsection{Comparison with kinetic theory}

In this subsection we evaluate $\bar{G}_{22}(X,p)$ using the
self-energies calculated in the previous subsection. Since both the
gain \eqref{eq:G22gainPi} and loss \eqref{eq:G22lossPi} terms are
proportional to $\delta(p^2)$, quasi-particle picture applies;
therefore, as a comparison we also calculate the distribution function
at $O(\alpha_s^2)$ by performing a perturbative solution of the
Boltzmann equation.

\subsubsection{Results from the above approximation}
\label{sec:results}

Inserting (\ref{eq:Pi11}) and (\ref{eq:Pi21}) into
(\ref{eq:G22gainPi}) and (\ref{eq:G22lossPi}) gives
\begin{align}
  \bar G_{22}^{gain}(X,p)=&\int_0^{X^+} dZ^+ \int_0^{X^-} dZ^-\frac{\pi}{p_\perp \tau_{X-Z}}\delta(p^2)\delta(y-\eta_{X-Z})\nonumber\\
  &\times\frac{1}{2}\int_{p_1,p_2,p_3} (2\pi)^4\delta(p+p_1-p_2-p_3)\overline{|M|^2} f^{cl}(Z,p_2)f^{cl}(Z,p_3),  \label{gain_init} \\
  \bar G_{22}^{loss}(X,p)=&-\int_0^{X^+} dZ^+ \int_0^{X^-} dZ^-\frac{\pi}{p_\perp \tau_{X-Z}}\delta(p^2)\delta(y-\eta_{X-Z})\nonumber\\
  &\times\frac{1}{2}\int_{p_1,p_2,p_3} (2\pi)^4\delta(p+p_1-p_2-p_3)
  f^{cl}(Z,p_1)f^{cl}\left( \frac{X+Z}{2} ,p \right). \label{loss_init} 
\end{align}
Both terms give a boost-invariant (rapidity-independent) particle
distribution. This is more transparent if one uses the following
variables
\begin{align}
  \tau_Z=\sqrt{2 Z^+ Z^-}, \qquad
  \eta_Z=\frac{1}{2}\ln\frac{Z^+}{Z^-}.
\end{align}
Since our approximation should break down at early times, we require
$\tau_Z>\tau_0$ with $\tau_0$ some initial time.  Since
$f^{cl}(X,p)\propto \delta(\eta-y)$, it is convenient to take
\begin{align}
  \int_{p_i}=\int\frac{d^4p_i}{(2\pi)^4}\theta(p_i^0)2\pi
  \delta(p_i^2)=\frac{1}{2}\int\frac{d y_i \, d^2 p_i}{(2\pi)^3}.
\end{align}

Let us start with the gain term. Using \eq{eq:fcl} we write
\begin{align}
  &\bar G_{22}^{gain}(X,p)=\frac{\delta(p^2)}{16p_\perp}\prod\limits_{i=1}^3\int\frac{d^2 p_i}{(2\pi)^2}(2\pi)^2\delta(\underline{p}+\underline{p}_1-\underline{p}_2-\underline{p}_3)\nonumber\\
  &\qquad\times f_\perp^{cl}(p_{2\perp})f_\perp^{cl}(p_{3\perp}) \int\limits_{\tau_0}^\tau \frac{d\tau_Z}{\tau_Z} \int\limits_{-\ln\frac{\tau}{\tau_Z}+\eta}^{\ln\frac{\tau}{\tau_Z}+\eta} d\eta_Z\frac{1}{\tau_{X-Z}}\delta(y-\eta_{X-Z}) \, \overline{|M|^2}\nonumber\\
  &\qquad\times\int d y_1 \delta(p_\perp e^{y}+p_{1\perp}
  e^{y_1}-p_{23\perp}e^{\eta_Z})\delta(p_\perp e^{-y}+p_{1\perp}
  e^{-y_1}-p_{23\perp}e^{-\eta_Z})\label{eq:G22gaintau}
\end{align}
with $p_{23\perp}\equiv p_{2\perp}+p_{3\perp}$.
By integrating out $y_1$ and $\eta_Z$ we have
\begin{align}
  &\bar G_{22}^{gain}(X,p)=\frac{\delta(p^2)}{16p_\perp}\prod\limits_{i=1}^3\int\frac{d^2{p}_i}{(2\pi)^2}(2\pi)^2\delta(\underline{p}+\underline{p}_1-\underline{p}_2-\underline{p}_3)\frac{f_\perp^{cl}(p_{\perp2})f_\perp^{cl}(p_{\perp3})}{\mathcal{P}^2(p_\perp,p_{1\perp}, p_{2\perp}, p_{3\perp})}\nonumber\\
  &\times\sum_{\{\eta_Z,y_1\}}\int_{\tau_0}^\tau
  \frac{d\tau_Z}{\tau_Z}\frac{1}{\tau_{X-Z}}\delta(y-\eta_{X-Z}) \,
  \overline{|M|^2} \,
  \theta\left(e^{\eta-\eta_Z}-\frac{\tau_Z}{\tau}\right)\theta\left(e^{\eta_Z-\eta}-\frac{\tau_Z}{\tau}\right), \label{eq:G22gaintau2}
\end{align}
where
\begin{align}
  &\mathcal{P}^2(p_\perp,p_{1\perp}, p_{2\perp}, p_{3\perp})\equiv \sqrt{[(p_{1\bot }-p_{\bot })^2-p_{23\bot }^2][(p_{1\bot }+p_{\bot })^2-p_{23\bot }^2]}
\end{align}
and the sum over $\{y_1, \eta_Z\}$ goes over the following two values
for each variable
\begin{align}
  &e^{y_1-y}= \frac{ \pm \mathcal{P}^2\left(p_{\bot },p_{1\bot },p_{2\bot },p_{3\bot }\right)-p_{1\bot }^2 + p_{23\bot}^2-p_{\bot }^2}{2 p_{1\bot } p_{\bot }},\nonumber\\
  &e^{\eta _Z-y}= \frac{\pm \mathcal{P}^2\left(p_{\bot },p_{1\bot
      },p_{2\bot },p_{3\bot }\right)-p_{1\bot }^2 +
    p_{23\bot}^2+p_{\bot }^2}{2 \, p_{23\bot } \, p_{\bot
    }}. \label{eq:eZy101}
\end{align}
The above two solutions for $\{y_1,\eta_Z\}$ respectively give
\begin{align}
  \sinh(y-\eta_Z)=\pm\frac{\mathcal{P}^2(p_\perp,p_{1\perp},
    p_{2\perp}, p_{3\perp})}{2 \, p_{23\perp} \,
    p_\perp}\equiv\pm\frac{\hat{p}_z}{p_\perp}.
\end{align}
Now let us integrate out $\tau_Z$. The $\delta$-function in
\eq{eq:G22gaintau2} gives
\begin{align}
  \tau _Z= \frac{\sinh (y-\eta )}{\sinh (y-\eta_Z)}\tau,\qquad
  \tau_{X-Z}=\frac{\sinh (\eta-\eta_Z )}{\sinh (y-\eta_Z)}\tau,
\end{align}
and the Jacobian
\begin{align}
  J\equiv\frac{1}{\left|\frac{d}{d\tau_Z}(y-\eta_{X-Z})\right|}=\frac{\tau_{Z}
    \, \tau_{X-Z}}{\tau |\sinh (y-\eta)|}.
\end{align}
From the above equations, we finally obtain
\begin{align}
  \bar G_{22}^{gain}(X,p)&=\frac{\delta(p^2)}{16p_\perp \, \tau}\prod\limits_{i=1}^3\int\frac{d^2\underline{p}_i}{(2\pi)^2}(2\pi)^2\delta(\underline{p}+\underline{p}_1-\underline{p}_2-\underline{p}_3)\frac{f_\perp^{cl}(p_{\perp2})f_\perp^{cl}(p_{\perp3})}{\mathcal{P}^2(p_\perp,p_{1\perp}, p_{2\perp}, p_{3\perp})}\nonumber\\
  &\times\frac{\overline{|M|^2}}{|\sinh(y-\eta )|}\theta\left(p_\perp|\sinh  (y-\eta )|-\hat{p}_z \tau_0/\tau\right)\theta\left(\hat{p}_z-p_\perp|\sinh  (y-\eta )|\right)\nonumber\\
  &\times\theta\left(e^{\eta-\eta_Z}- \frac{p_\perp|\sinh (y-\eta
      )|}{\hat{p}_z}\right)\theta\left(e^{\eta_Z-\eta}-
    \frac{p_\perp|\sinh (y-\eta )|}{\hat{p}_z}\right), \label{gain_final}
\end{align}
where $\eta-\eta_Z$ should have the same sign as $y-\eta_Z$ and
$y-\eta$, and $\{y_1,\eta_Z\}$ only assume the values in
(\ref{eq:eZy101}) which ensure that $\tau_Z$ is positive.

For the loss term \eqref{loss_init} evaluation appears to be more
complicated in general. The difficulty is in the $(X+Z)/2$ in the
argument of one of the $f^{cl}$ in \eqref{loss_init}. It appears that
to obtain something similar to kinetic theory one has to replace
\begin{align}
  \label{eq:repl}
  \frac{X+Z}{2} \to Z
\end{align}
in the argument of $f^{cl}$ in \eqref{loss_init}. This is an {\sl ad
  hoc} assumption, particularly for the plus and minus components
$X^\pm, Z^\pm$, which does not follow from the orderings (i) and (ii)
considered above. In fact, it can never be realized in the case of
ordering (ii). Within ordering (i) one could imagine a situation where
\begin{align}
  \label{eq:iii}
  Z^+ \gg X^+ - Z^+ \gg 1/Q_s
\end{align}
and the replacement \eqref{eq:repl} may be justified. Such a condition
is a further refinement of the ordering (ii) and was not needed for
the gain term. Below we will assume that the ordering \eqref{eq:iii}
applies and make the substitution \eqref{eq:repl}.

Working in this approximation one can easily integrate out $\eta_Z$
and $y_1$ in \eq{loss_init} and get
\begin{align}
  &\bar G_{22}^{loss}(X,p)=-\frac{\delta(p^2)}{16p_\perp}\prod\limits_{i=1}^3\int\frac{d^2\underline{p}_i}{(2\pi)^2}(2\pi)^2\delta(\underline{p}+\underline{p}_1-\underline{p}_2-\underline{p}_3)\nonumber\\
  &\qquad\times f_\perp^{cl}(p_\perp)f_\perp^{cl}(p_{1\perp}) \int_{\tau_0}^\tau \frac{d\tau_Z}{\tau_Z} \frac{1}{\tau_{X-Z}}\delta(y-\eta_{X-Z})\overline{|M|^2}\nonumber\\
  &\qquad\times\int d y_2 d y_3\delta((p_\perp+p_{1\perp}) e^{y}-p_{2\perp}e^{y_2}-p_{3\perp}e^{y_3})\nonumber\\
  &\times\delta((p_\perp+p_{1\perp})
  e^{-y}-p_{2\perp}e^{-y_2}-p_{3\perp}e^{-y_3})\theta\left(e^{\eta-y}-\frac{\tau_Z}{\tau}\right)\theta\left(e^{y-\eta}-\frac{\tau_Z}{\tau}\right)\label{eq:G22lossl2}
\end{align}
with $\eta_Z=y$. Now we are left with
\begin{align}
  \delta(y-\eta_{X-Z})=\frac{\tau-\tau_Z}{\tau}\delta(y-\eta).
\end{align}
At the end, we have
\begin{align}
  &\bar G_{22}^{loss}(X,p)=-\frac{\delta(p^2)}{16p_\perp \tau}\ln\left(\frac{\tau}{\tau_0}\right)\prod\limits_{i=1}^3\int\frac{d^2\underline{p}_i}{(2\pi)^2}(2\pi)^2\delta(\underline{p}+\underline{p}_1-\underline{p}_2-\underline{p}_3)\nonumber\\
  &\qquad\times
  \delta(y-\eta)\frac{f_\perp^{cl}(p_\perp)f_\perp^{cl}(p_{1\perp})}{\mathcal{P}^2(p_{2\perp},
    p_{3\perp},p_\perp,p_{1\perp})}\sum\limits_{\{y_2,y_3\}}\overline{|M|^2} \label{loss_final}
\end{align}
with $\{y_2,y_3\}$ summed over two values each given by the solutions
to the equations resulting from the last two $\delta$ functions in
(\ref{eq:G22lossl2}). 

We are now going to compare the results \eqref{gain_final} and
\eqref{loss_final} of this Subsection for the gain and loss terms with
the predictions of kinetic theory.


\subsubsection{Results from the Boltzmann equation}

In the boost-invariant and dilute ($f \ll 1$) system, the Boltzmann
equation (\ref{eq:boltEqua}) reduces to
\cite{Mueller:1999pi}\footnote{For completeness we have included the
  standard derivation of the Boltzmann equation for gluons in
  Appendix~\ref{B}.}
\begin{align}
  &\left(\frac{\partial}{\partial t} - \frac{1}{t} p_z
    \frac{\partial}{\partial p_z}\right)
  f= C[f](t,p_\perp,p_z)\nonumber\\
  &\qquad\qquad\equiv\frac{1}{4
    \omega_p}\int_{p_1,p_2,p_3}\overline{|M|^2}[f_2f_3-f f_1](2\pi)^4
  \delta(p+p_1-p_2-p_3).
\end{align}
While the system we consider is boost-invariant (rapidity
independent), we will concentrate on central rapidity, $\eta =0$,
throughout this Subsection. At $\eta =0$ one has $\tau = t$.  

In order to see the connection to the calculation in our formalism, we
write
\begin{align}
  f=\sum\limits_{n=0}^\infty f^{(n)} \qquad\text{with $f^{(n)}$ being
    of $O(\alpha_s^{2n})$}.
\end{align}
The terms $f^{(n)}$ are to be found from solving the Boltzmann
equation order-by-order in the coupling $\as$. The initial condition
(the value of $f$ before the collision term becomes important) is
given by saturation dynamics (cf. \eq{eq:fcl}),
\begin{align}\label{init_cond}
  f^{(0)}(t,p)=\frac{1}{t} f_\perp(p_\perp) \delta(y),
\end{align}
which satisfies the ``free" Boltzmann equation 
\begin{align}
  \left(\frac{\partial}{\partial t} - \frac{1}{t} p_z
    \frac{\partial}{\partial p_z}\right) f^{(0)}=0.
\end{align}
The higher orders in $\as$ can be calculated by iteration
\begin{align}\label{eq:fn1}
  f^{(n+1)}=\int_{t_0}^t dt_1 \, C[f^{(n)}](t_1,t p_z/t_1)+C^{(n+1)}(t
  p_z/t_0)
\end{align}
with $C^{(n+1)}$ the constant of integration at $O(\alpha_s^{2(n+1)})$
and $t_0$ some initial time when the Boltzmann dynamics starts to
apply (e.g. $t_0 \sim 1/Q_s$).

For comparison with the results of the previous Subsection we need
only to evaluate terms proportional to $O(\alpha_s^2)$. According to
(\ref{eq:fn1}), we first evaluate
\begin{align}
  & C[f^{(0)}](t,p_\perp,p_z) = \frac{1}{32\pi t^2}\prod\limits_{i=1}^3 \int \frac{d^2\underline{p}_i}{(2\pi)^2} (2\pi)^2 \delta(\underline{p}+\underline{p}_1-\underline{p}_2-\underline{p}_3) \nonumber\\
  &\times\left.\left[\frac{f_\perp(p_{2\perp}) f_\perp(p_{3\perp})}{2
        \omega_p \omega_{p_1}
      }  \, \overline{|M|^2} \, \delta(\omega+\omega_1- p_{2\perp}- p_{3\perp})\right.\right|_{p_{1z}=-p_z, \, p_{2z} = p_{3z}=0} \nonumber \\
  &\left.\left. -\delta(y) \, \frac{2 \, f_\perp(p_\perp) \,
        f_\perp(p_{1\perp})}{p_\perp \mathcal{P}^2(p_{2\perp},
        p_{3\perp},p_\perp,p_{1\perp})} \, \overline{|M|^2}
    \right|_{p_{1z}=0, \, |p_{2z}| = |p_{3z}| =
      \frac{\mathcal{P}^2(p_{2\perp},
        p_{3\perp},p_\perp,p_{1\perp})}{2 (p_\perp + p_{1\perp})}}
  \right]. \label{Cf1}
\end{align}
To satisfy the initial conditions at time $t = t_0$ we put the
integration constant to zero, $C^{(1)}=0$. Substituting \eq{Cf1} into
\eq{eq:fn1} and integrating yields
\begin{align}\label{eq:f1}
  & f^{(1)} = \frac{1}{32\pi t} \, \prod\limits_{i=1}^3\int\frac{d^2\underline{p}_i}{(2\pi)^2} \, (2\pi)^2 \, \delta(\underline{p}+\underline{p}_1-\underline{p}_2-\underline{p}_3) \\
  &\times \left. \left[\frac{f_\perp(p_{2\perp})
        f_\perp(p_{3\perp})}{|p_z|\mathcal{P}^2(p_\perp,p_{1\perp},p_{2\perp},
        p_{3\perp})
      } \, \theta(|p_z|-\hat{p}_z t_0/t) \, \theta(\hat{p}_z-|p_z|) \, \overline{|M|^2}\right|_{p_{1z}=-p_z, \, p_{2z} = p_{3z}=0} \right.\nonumber\\
  &\left. \left. -\delta(y)\ln\left(\frac{t}{t_0}\right) \frac{2 \,
        f_\perp(p_\perp)f_\perp(p_{1\perp})}{p_\perp
        \mathcal{P}^2(p_{2\perp}, p_{3\perp},p_\perp,p_{1\perp})} \,
      \overline{|M|^2} \right|_{p_{1z}=0, \, |p_{2z}| = |p_{3z}| =
      \frac{\mathcal{P}^2(p_{2\perp},
        p_{3\perp},p_\perp,p_{1\perp})}{2 (p_\perp +
        p_{1\perp})}}\right]. \notag
\end{align}
We obtain a linear combination of $\sim 1/t$ and $\sim \delta(y) \,
(1/t) \, \ln (t/t_0)$ terms. This is exactly the same $t$ and
$y$-dependence as that in the previous Subsection for the gain and
loss terms respectively (if we apply $\eta=0$ to those results).

Knowing the distribution function $f$ one can calculate the
energy-momentum tensor using
\begin{align}
  \label{eq:EMT}
  T^{\mu\nu} (x) = 2\, \int\limits_p \, p^\mu \, p^\nu \, f (x,p).
\end{align}
Clearly, the initial conditions \eqref{init_cond}, or, equivalently,
the classical gluon correlator \eqref{eq:fcl} give $T^{\mu\nu} \sim
1/\tau$ for all the non-zero components of the energy-momentum tensor,
along with the longitudinal pressure $P_L = T^{33} (\eta=0) = 0$: this
behavior corresponds to free streaming of gluons. Adding the
correction $f^{(1)}$ from \eq{eq:f1} we obtain
\begin{subequations}\label{eq:results}
\begin{align}
  & \epsilon = \epsilon^{(0)} + \epsilon^{(1)} = \frac{A^{(0)} + \as^2 \, A^{(1)}}{\tau} - \frac{\as^2 \, B^{(1)}}{\tau} \, \ln \frac{\tau}{\tau_0}, \\
  & P_T = P_T^{(0)} + P_T^{(1)} = \frac{A^{(0)} + \as^2 \, A^{(1)} - \as^2 \, B^{(1)}}{\tau} - \frac{\as^2 \, B^{(1)}}{\tau} \, \ln \frac{\tau}{\tau_0}, \\
  & P_L = P_L^{(1)} = \frac{\as^2 \, B^{(1)}}{\tau} .
\end{align}
\end{subequations}
The exact values of the coefficients $A^{(0)}, A^{(1)}$ and $B^{(1)}$
can be found by explicit integration: their exact values are not
important to us, as long as $A^{(1)}$ and/or $B^{(1)}$ are not
zero. The $\sim 1/\tau$ and $\sim (1/\tau) \, \ln (\tau/\tau_0)$ terms
that $A^{(1)}$ and $B^{(1)}$ multiply constitute a deviation from the
$\sim 1/\tau$ free-streaming behavior of the classical gluon
fields. We conclude that kinetic theory predicts a deviation from free
streaming after including a single $2 \to 2$ rescattering correction
to the classical gluon correlator. This prediction appears to agree
with the results of the approximate calculations carried out above,
after certain approximations were made. We will verify this prediction
in \cite{KovchegovWu}.


\section{Free Streaming}
\label{sec:free}

Here we show how the above calculation can lead to different results
depending on how the large-$X^+ - Z^+$, $X^- - Z^-$ is imposed. In
particular, we demonstrate that ordering (ii) from above does not lead
to kinetic theory, but rather to free streaming of the produced
gluons.


\subsection{Free streaming after rescattering}

In the calculations of Sec.~\ref{sec:results} the integrations over
$\tau_Z$ extend all the way up to $\tau$, in an apparent violation of
the large-$X^+ - Z^+$, $X^- - Z^-$ assumption used in deriving
Eqs.~\eqref{gain_init} and \eqref{loss_init}. To preserve the main
results \eqref{gain_final} and \eqref{loss_final} of
Sec.~\ref{sec:results} while satisfying the large-$X^+ - Z^+$, $X^- -
Z^-$ condition one could impose the ordering (i) from
Sec.~\ref{sec:Gresc} in the following way:
\begin{align}\label{cond(i)}
\mbox{(i)} \ \   \tau_0 \ll \tau_z \ll \xi \, \tau \ll \tau
\end{align}
with a small parameter $\xi \ll 1$ (but not too small, such that $\xi
\tau \gg \tau_0$ still). For instance, we may have $\xi = \as^\lambda$
with some positive power $\lambda >0$. Replacing $\tau \to \xi \tau$
in the integration limits of \eq{eq:G22gaintau} would still lead to
$\bar G_{22}^{gain}(X,p) \sim 1/\tau$, just like in
\eq{gain_final}. Similarly, replacing $\tau \to \xi \tau$ in the upper
integration limit and in the theta-functions of \eq{eq:G22lossl2} one
still obtains $\bar G_{22}^{loss}(X,p) \sim (1/\tau) \ln(\tau/\tau_0)
\, \delta (y-\eta)$, just like in \eqref{loss_final}. While the
prefactors may be modified by the $\tau \to \xi \tau$ substitution,
the $\tau$ and $\eta$ dependence of the gain and loss terms would
remain the same. Hence, \eq{cond(i)} appears to provide a more proper
way of imposing the condition (i) on the calculation in
Sec.~\ref{sec:results}.

It appears natural that in addition to the ordering \eqref{cond(i)}
one also considers
\begin{align}\label{cond(ii)}
  \mbox{(ii)} \ \ \tau_0 \ll \tau_z \ll \zeta \, \tau_0 \ll \tau
\end{align}
where $\zeta \gg 1$ is a large parameter. For $\tau_0 = 1/Q_s$ one may
have $\zeta = 1/\as^\omega$ with $\omega >0$ such that $\zeta \tau_0 =
1/(\as^\omega \, Q_s)$. While $\zeta$ is large, it should not be too
large, such that $\zeta \, \tau_0 \ll \tau$ still. \eq{cond(ii)} is
consistent with the condition (ii) from Sec.~\ref{sec:Gresc} and also
provides a way to impose the large-$X^+ - Z^+$, $X^- - Z^-$ condition.

Replacing the upper limit of the $\tau_Z$ integration in
\eq{eq:G22gaintau2} by using $\tau \to \zeta \, \tau_0$ (along with
the same replacement in the arguments of $\theta$-functions, which we
discard below since after for small $\tau_Z$ they are automatically
satisfied) one gets
\begin{align}\label{gain_free}
  \bar G_{22}^{gain}(X,p) \sim \int\limits_{\tau_0}^{\zeta \, \tau_0}
  \frac{d\tau_Z}{\tau_Z}\frac{1}{\tau_{X-Z}}\delta(y-\eta_{X-Z})
  \approx \int\limits_{\tau_0}^{\zeta \, \tau_0}
  \frac{d\tau_Z}{\tau_Z}\, \frac{1}{\tau}\delta(y-\eta) =
  \frac{1}{\tau}\delta(y-\eta) \, \ln \zeta.
\end{align}
We have employed $\tau_z \ll \zeta \, \tau_0 \ll \tau$ condition in
simplifying \eq{gain_free}. (Strictly-speaking we have assumed a
somewhat stronger condition $Z^\pm \ll \zeta \, Z_0^\pm \ll X^\pm$
consistent with the original ordering (ii) to approximate $\tau_{X-Z}
\approx \tau$ and $\eta_{X-Z} \approx \eta$.) We see that with the
ordering (ii), \eqref{cond(ii)}, the gain contribution to the
correlation function still scales as $1/\tau$, but now it is
multiplied by $\delta(y-\eta)$: the $\delta$-function leads to zero
longitudinal pressure, making this $\bar G_{22}^{gain}(X,p)$
consistent with free streaming.

The loss term is treated similarly: performing the $\tau \to \zeta \,
\tau_0$ replacement in \eq{eq:G22lossl2} gives
\begin{align}\label{loss_free}
  \bar G_{22}^{loss}(X,p) \sim \int\limits_{\tau_0}^{\zeta \, \tau_0}
  \frac{d\tau_Z}{\tau_Z}\frac{1}{\tau_{X-Z}}\delta(y-\eta_{X-Z})
  \approx \int\limits_{\tau_0}^{\zeta \, \tau_0}
  \frac{d\tau_Z}{\tau_Z}\, \frac{1}{\tau}\delta(y-\eta) =
  \frac{1}{\tau}\delta(y-\eta) \, \ln \zeta
\end{align}
which is also consistent with free streaming.\footnote{Note that due
  to the {\sl ad hoc} approximation \eqref{eq:repl} made in evaluating
  the loss term above, its late-time asymptotics should be derived by
  evaluating the diagrams in the second row of \fig{fig:g22lossgain}
  from scratch: in \cite{KovchegovWu} this will be done in the
  framework of the $\varphi^4$ theory.}

We conclude that while the calculations of Sec.~\ref{sec:results}
appear to be consistent with kinetic theory if the ordering (i) is
imposed via \eqref{cond(i)}, the same calculations are consistent with
free streaming if the ordering (ii) is imposed with the help of
\eqref{cond(ii)}. Therefore, since at this level of calculational
precision we can not say whether the ordering (i) or (ii) is correct,
we can not tell whether our calculation supports kinetic theory or the
free-streaming scenario advocated in \cite{Kovchegov:2005ss}.


\subsection{A general argument for free streaming} 

For completeness, let us briefly recap the free-streaming argument
from \cite{Kovchegov:2005ss}, but now for the correlation function
$\bar G_{22}$ considered in this work. (In \cite{Kovchegov:2005ss} the
argument was applied to the energy-momentum tensor of the medium
produced in heavy ion collisions.)  In a general case, involving all
the possible multiple interactions and rescatterings, one can still
write the $\bar G_{22}$ correlation function as a sum of the three
diagrams in the top row of \fig{fig:Pi11}, but now without requiring
that $\Pi_{ij}$ include 2PI diagrams only. The circles now denote any
(connected) diagram. The resulting expression is the same as in
\eq{eq:G22formal}:
\begin{align}\label{G1}
  \bar G_{22} (k, k')= - i
  (2\pi)^2\delta(\underline{k}+\underline{k}') & \left[ G_R(k) {\bar
      \Pi}_{11} (k, k') G_R (k') + G_R (k) {\bar \Pi}_{12} (k, k')
    G_{12}^{(0)} (k') \right. \notag \\ & \left.  + G_{22}^{(0)} (k)
    {\bar \Pi}_{21} (k, k') G_R(k') \right] .
\end{align}
Since all ${\bar \Pi}_{ij}$ are Lorentz-invariant, they can only be
functions of $k^2$, $k^{\prime \, 2}$ and $k \cdot k'$. The ${\bar
  \Pi}_{ij}$'s can also be functions of $p \cdot k \sim k^-$, $p'
\cdot k \sim k^+$, $p \cdot k' \sim k^{\prime \, -}$, and $p' \cdot k'
\sim k^{\prime \, +}$ with $p$ and $p'$ the momenta of the nucleons in
nucleus $A_1$ and $A_2$. In \cite{Kovchegov:2005ss} a ``dictionary"
was established by going through a number of examples. According to
this ``dictionary" the Fourier transform of the correlation function
into coordinate space,
\begin{align}
  \bar G_{22} (x_1 , x_2) = \int \frac{d^4 k}{(2 \pi)^4} \frac{d^4
    k'}{(2 \pi)^4} \, e^{- i k \cdot x_1 - i k' \cdot x_2} \, \bar
  G_{22} (k, k'),
\end{align}
converts (modulo some prefactors and an index shift of the Bessel
function resulting from the transform)
\begin{subequations}
\begin{align}
  k^2, k \cdot k', k^{\prime \, 2} \, & \longrightarrow \,
  \frac{2 k_T}{\tau}, \label{k20} \\
  k^\pm \, & \longrightarrow \, \frac{k_T \, e^{\pm
      \eta}}{\sqrt{2}} \label{kplus}
\end{align}
\end{subequations}
with $\eta$ the space-time rapidity. According to \eq{k20}, the
leading late-$\tau$ contribution comes from putting $k^2 = k \cdot k'
= k^{\prime \, 2} =0$ in the ${\bar \Pi}_{ij}$'s, with corrections to
it being suppressed by powers of $1/\tau$.

Let us illustrate this using the ${\bar \Pi}_{11}$ term in
\eq{G1}. Putting $k^2 = k \cdot k' = k'^{2} =0$ in ${\bar \Pi}_{11}$
and integrating that term in \eqref{G1} over $k^-$ and $k^{\prime \,
  -}$ yields
\begin{align}\label{G2}
  & {\bar G}_{22} (x_1, x_2) = -i \, \theta (x_1^+) \, \theta (x_2^+)
  \, \int \frac{d^2 k_{\perp}}{(2\pi)^2} e^{i {\un k} \cdot ({\un x}_1
    - {\un x}_2)} \int\limits_{-\infty}^\infty \frac{d k^+}{4\pi (k^+
    + i \epsilon)} \\ &\times \, \int\limits_{-\infty}^\infty \frac{d
    k'^+}{4\pi (k'^+ + i \epsilon)} e^{-i k^+ x_1^- - i
    \frac{k_\perp^2}{2 (k^+ + i \epsilon)} x_1^+ - i k'^+ x_2^- - i
    \frac{k_\perp^2}{2 (k'^+ + i \epsilon)} x_2^+} \, {\bar \Pi}_{11}
  (k,k') \Bigg|_{k^2 = k \cdot k' = k'^{2} =0, \, \un{k}' = -
    \un{k}}. \notag
\end{align}
In arriving at \eq{G2} we have also neglected the $k^\pm, k'^\pm$
dependence in ${\bar \Pi}_{11}$: below we will briefly outline how
this can be reinstated. Employing
\begin{align}
  \int\limits_{-\infty}^\infty \frac{d k^+}{4\pi (k^+ + i \epsilon)}
  \, e^{-i k^+ x^- - i \frac{k_\perp^2}{2 (k^+ + i \epsilon)} x^+} = -
  \frac{i}{2} \theta (x^-) \, J_0 (k_\perp \tau) \label{Int1}
\end{align}
we obtain
\begin{align}\label{G4}
  \bar G_{22} (x_1, x_2) = & \, \frac{i}{4} \theta (x_1^+) \, \theta
  (x_2^+) \, \theta (x_1^-) \, \theta (x_2^-) \, \int \frac{d^2
    k_{\perp}}{(2\pi)^2} e^{i {\un k} \cdot ({\un x}_1 - {\un x}_2)}
  \notag \\ & \times \, J_0 (k_\perp \tau_1) J_0 (k_\perp \tau_2) \,
  {\bar \Pi}_{11} (k,k') \Bigg|_{k^2 = k \cdot k' = k'^{2} =0, \,
    \un{k}' = - \un{k}}.
\end{align}
Clearly, 
\begin{align}\label{Gasym}
  \bar G_{22} (x_1, x_2) \Bigg|_{\tau_1 = \tau_2 = \tau \to \infty}
  \sim \frac{1}{\sqrt{\tau_1 \, \tau_2}} \Bigg|_{\tau_1 = \tau_2 =
    \tau} = \frac{1}{\tau}.
\end{align}
Hence the leading contribution to the correlator is $\sim 1/\tau$,
corresponding to free streaming, as the energy-momentum tensor that
one would obtain from the correlator \eqref{Gasym} would scale as
$T^{\mu\nu} \sim 1/\tau$. If there exist terms proportional to, say,
Bjorken hydrodynamics \cite{Bjorken:1982qr}, which has $T^{\mu\nu}
\sim 1/\tau^{4/3}$, they would be subleading compared to the
free-streaming term of \eq{Gasym}.

Including the $k^\pm, k'^\pm$-dependent terms in the above calculation
would only add space-time rapidity dependence in the correlation
function owing to \eq{kplus}, without changing the conclusion
\eqref{Gasym} about the late-time asymptotics. Finally, the argument
applies analogously to the ${\bar \Pi}_{12}$ and ${\bar \Pi}_{21}$
terms in \eq{G1}.

The remaining question is whether the leading asymptotics happens to
have a zero coefficient, that is, what if
\begin{align}
  {\bar \Pi}_{11} (k,k') + \frac{1}{2} \left[ {\bar
      \Pi}_{21} (k,k') - {\bar \Pi}_{12} (k,k') \right] \Bigg|_{k^2 =
    k \cdot k' = k'^{2} =0, \, \un{k}' = - \un{k}} \overset{?}{=} 0 \,
  ?
\end{align} 
In \cite{Kovchegov:2005ss} it was shown that for the energy-momentum
tensor this is not the case. The leading late-time contribution was
shown to be proportional to the particle (gluon) production cross
section. Specifically, the energy density was shown to be
\begin{align}\label{edens}
  \epsilon (\tau, \eta, \un{b}) \Bigg|_{\tau \to \infty} =
  \frac{1}{\tau} \int d^2 k_T \, k_T \, \frac{dN}{d^2 k_T \, d \eta \,
    d^2 b_\perp}.
\end{align}
Hence the leading free-streaming term is non-zero as long as one can
define the multiplicity of produced gluons $dN/d^2 k_T \, d \eta \,
d^2 b_\perp$, that is as long as perturbation theory
holds.\footnote{The distribution of produced gluons, $dN/d^2 k_T \, d
  \eta \, d^2 b_\perp$, requires an IR cutoff once collinear-divergent
  corrections are included: such cutoff cancels in the integral of
  \eq{edens} such that the energy density is independent of the cutoff
  (see \cite{Kovchegov:2007vf}).}


\section{Conclusions and Outlook}
\label{sec:disc}

In this paper, we adapted the Schwinger-Keldysh formalism to study
heavy-ion collisions in a perturbative QCD approach. We calculated the
gluon two-point correlation function $G_{22}^{a\mu,b\nu}$ at $O(g^{6}
A^{\frac{2}{3}})$ in the lowest-order classical approximation of the
MV model. We found that at large $\tau$ the (quasi-) particle picture
emerges from the classical field calculation at this order in the
sense that
\begin{align}
  G_{22}^{a\mu,b\nu}(X,p)\propto
  \frac{2\pi}{\tau}\delta^{ab}\sum\limits_{\lambda=\pm}\epsilon_\lambda^\mu(k_1)
  \epsilon_{\lambda}^{*\nu}(-k_2)\delta(p^2)\delta(y-\eta).\label{summary01}
\end{align}

Motivated by this observation, we evaluated a subset of diagrams at
$O(g^{16} A^{\frac{4}{3}})$, which are beyond the classical field
approximation, and corresponds to a $2\to 2$ rescattering of the
classically produced gluons. Each of these diagram includes two
sub-diagrams of $O(g^{6} A^{\frac{2}{3}})$, described by
$G_{22}^{a\mu,b\nu}$ in (\ref{summary01}) each. In our calculation the
rescattering occurs at some space-time point $Z^\mu$ while the gluon
distribution is measured at another point $X^\mu$. We made the
following approximations:
\begin{enumerate}
\item{$\tau_Z\equiv\sqrt{2Z^+Z^-} \gg 1/Q_s$}\\
  Under this assumption, each sub-diagram took the form in (\ref{summary01}). 
\item{$\tau_{X-Z}\equiv\sqrt{2(X^+-Z^+)(X^--Z^-)}\gg 1/Q_s$}\\
  Under this assumption, the gluons after the rescattering can be
  taken to be quasi-classical particles. That is, they travel along a
  classical trajectory $X^3-Z^3=(X^0-Z^0)p^3/p^0$ with $p^\mu$ the
  four-momentum of the gluons.
\end{enumerate}
Under the above approximations, which are equivalent to case (i)
listed above (or, for the loss term, by additionally imposing $\tau_Z
\gg \tau_{X-Z}$), we find the rescattering correction consistent with
a power series in $\as$ solution of the Boltzmann equation.

However, one needs to make a more detailed calculation to justify our
approximations above. In our approximations of
Eqs.~(\ref{eq:G22gaintau}) and (\ref{eq:G22lossl2}) we put the upper
limit of $\tau_Z$ integration to be $\tau$ in apparent violation of
the assumption 2. We also do not know {\sl a priori} whether our
assumption 2 correctly represents the full diagrammatic calculation,
since one may take another limit instead, $\tau\gg\tau_Z$,
corresponding to case (ii) by the above counting. In this case $\bar
G_{22}^{gain}$ and $\bar G_{22}^{loss}$ are still respectively given
by (\ref{eq:G22gaintau}) and (\ref{eq:G22lossl2}) with another upper
limit for $\tau_Z$ integration, as detailed in Sec.~\ref{sec:free}. If
one takes the limit $\tau\to \infty$ in this case, both the gain and
loss terms are proportional to $\delta(y-\eta)/\tau$. That is, after
the rescattering the gluons assume a distribution similar to that for
free-streaming particles. In the companion paper \cite{KovchegovWu} we
perform a detailed calculation in the framework of the $\lambda
\varphi^4$ theory to explicitly identify which approximation is
correct.


\acknowledgments

The authors would like to thank Mauricio Martinez for the extensive
discussions of thermalization in heavy ion collisions which got YK
interested in this project. We also thank Hong Zhang for useful
discussions and advice.  This material is based upon work supported by
the U.S. Department of Energy, Office of Science, Office of Nuclear
Physics under Award Number DE-SC0004286.


\appendix

\section{The classical field limit}
\label{sec:class}

\begin{figure}[tbp]
\begin{center}
\includegraphics[width=0.9\textwidth]{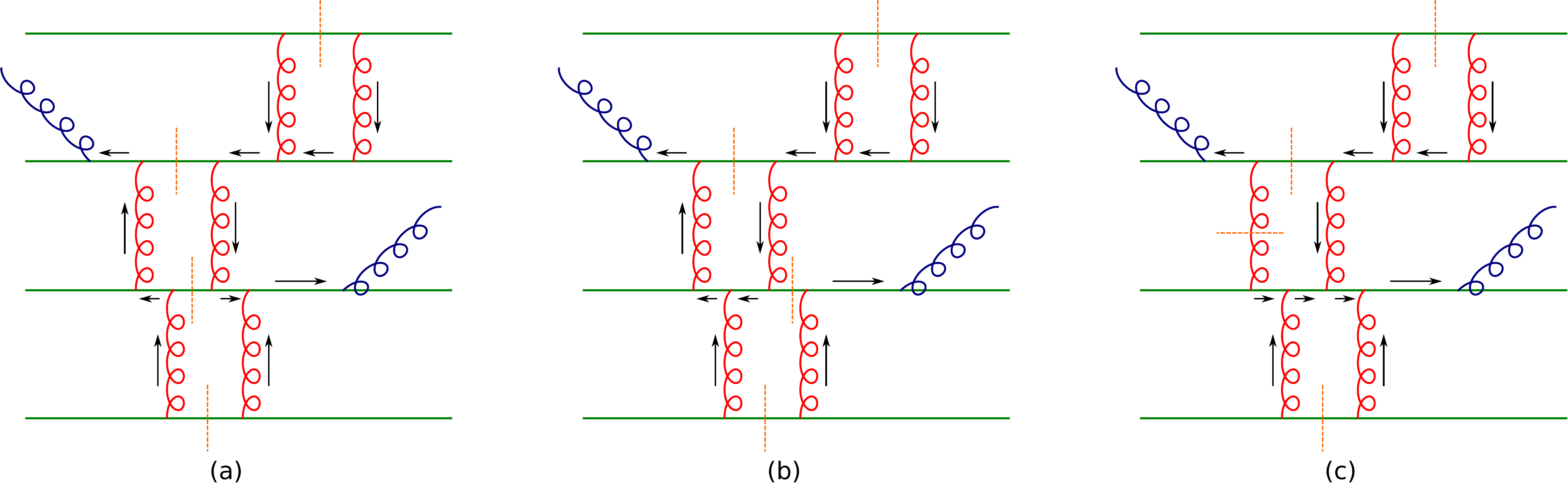}
\end{center}
\caption[*]{Illustration of diagrams in the classical limit. Diagrams
  (a) and (b) are separated into a product of two sub-diagrams
  connected to each other by $S^{(0)}_{22}$ cut quark
  propagators. Each of the sub-diagrams contributes to the classical
  field. In contrast, diagram (c) does not have such a
  separation. However, it is canceled in the eikonal
  approximation. Here, each orange dashed line crosses a quark $2-2$
  propagator indicating that the quark is on mass-shell and each
  retarded Green function is associated with an arrow pointing in the
  increasing time direction.}
\label{fig:AAeg}
\end{figure}

In this Appendix we will verify the classical field limit of our
formalism. The limit involves (a) making the eikonal approximation of
the quark lines from the nuclear wave functions; and (b) keeping only
the diagrams of order $\frac{1}{g^2} (g^4 A^{\frac{1}{3}})^{n_q}$ with
$n_q$ the number of the quark lines. Specifically, we shall show that
in this limit the gluon two-point function $G_{22}^{a\mu,b\nu}$
reduces to a product of classical gluon fields, i.e.,
\begin{align}
  G^{a\mu,b\nu}_{22}(x,y)=\left< A_{cl}^{a\mu}(x) A_{cl}^{b\nu}(y)
  \right>
  \label{eq:GclassLimit}
\end{align}
with the angle brackets denoting the averaging from \eq{eq:DMinit3}.
As illustrated in Figs.~\ref{fig:AAeg}(a) and \ref{fig:AAeg}(b), we
shall prove that any diagram of order $\frac{1}{g^2} (g^4
A^{\frac{1}{3}})^{n_q}$ that does not vanish in the eikonal
approximation is separated into two disconnected pieces by the set of
all $S^{(0)}_{22}$ (one on each quark line) in the eikonal
approximation. Each piece is connected by retarded Green function and
hence contributes to the classical field.


\subsection{The eikonal approximation of the quark lines}

At high energies, the recoil of the valence quarks from radiating soft
gluons is negligible. In this case one can make the so-called eikonal
approximation to the quark lines
\cite{Mueller:1989st,McLerran:1993ni,McLerran:1994vd,Kovchegov:1996ty}. In
each diagram for our problem, the valence quarks of nucleus 1 receive
some momentum transfer, $l$, from the scattering with other
partons. $l$ is typically much softer than $P_1^+$. This allows us to
approximate the free quark propagator by
\begin{align}
  S^{(0)}_{ij} (P_1+l)\approx \gamma^- \, \delta_{ij} \,
  \left(\begin{array}{cc}
      0&\frac{i}{l^--i\epsilon}\\
      \frac{i}{l^-+i\epsilon}&\pi\delta(l^-)
\end{array}\right).\label{eq:S1}
\end{align}  
Here we ignore the quark mass. Similarly, for a valence quark of
nucleus 2 with momentum transfer $l$, one has
\begin{align}
  S^{(0)}_{ij} (P_2+l)\approx \frac{\slashed{P}_2+\slashed{l}}{2P_2^-}
  \, \delta_{ij} \left(\begin{array}{cc}
      0&\frac{i}{l^+-i\epsilon}\\
      \frac{i}{l^++i\epsilon}&\pi\delta(l^+)
\end{array}\right).\label{eq:S2}
\end{align}
Here, we have kept $\slashed{l}$ because it may not always give a
suppressed contribution compared to $\slashed{P}_2$ in the $A^+=0$
light-cone gauge.

In this paper we shall not consider the small-$x$ quark
production. That is, all the quark lines come from the two nuclear
wave functions. In this case the eikonal approximation only involves
replacing the quark propagator in (\ref{eq:Gq}) by Eqs. (\ref{eq:S1})
and (\ref{eq:S2}). Below we shall show that substituting these two
equations for each quark propagator $S$ in the diagrams of order
$\frac{1}{g^2} (g^4 A^{\frac{1}{3}})^{n_q}$ for $G_{22}^{a\mu,b\nu}$
separates each diagram into two sub-diagrams connected to each other
only by the $S^{(0)}_{22}$ (cut) quark propagators.


\subsection{Diagrams in the classical field approximation}

Let us focus on a generic diagram with $n_q$ valence quark
lines. First of all, the diagram should be connected. Otherwise, it
can be separated into the product of connected sub-diagrams. Among
these sub-diagrams, there must exist at least one diagram either
without any gluon radiation or with one gluon radiation. Due to
unitarity, connected diagrams without radiation cancel. And diagrams
with one radiated gluons also vanish after the average over the
initial distribution in (\ref{eq:DMinit3}) due to the color neutrality
of the two-nuclei source. Therefore, it has to be connected.

Second of all, the diagram should be of order $g^{4n_q-2}
A^{\frac{n_q}{3}}$ in order to give a non-vanishing contribution in
the classical limit. The factor
$A^{\frac{n_q}{3}}\propto\left(\frac{A}{S_\perp}\right)^{n_q}$ results
from the average at the initial time in \eq{eq:DMinit3}. We shall
prove that the diagram gives a non-vanishing contribution to
$G_{22}^{a\mu,b\nu}$ in the eikonal approximation only if it has a
$S^{(0)}_{22}$ on each quark line. An example with $n_q=4$ is shown in
Fig.~\ref{fig:AAeg}. We will show that the first two diagrams, each of
which is separated into two sub-diagrams connected by $S^{(0)}_{22}$
on each quark line, give non-vanishing contributions while the third
one vanishes in the eikonal approximation. Including all the
non-vanishing diagrams such as those in Figs.~\ref{fig:AAeg} (a) and
(b) leads to the classical field approximation in
(\ref{eq:GclassLimit}).

In order to prove the above statements, we need first to count the
number of 2-2 propagators in the diagram. Let us assume that there are
$n_{4}^{cl}$ four-gluon vertices with one ``$1$'' field, $n_{4}^{qu}$
4-gluon vertices with three ``$1$'' fields, $n_{3}^{cl}$ three-parton
vertices with one ``1" field and $n_{3}^{qu}$ three-parton vertices
with three ``1" fields. Then, the diagram is of order $g^{n_D}$ with
\begin{align}
  n_D=2(n_{4}^{cl}+ n_{4}^{qu})+(n_{3}^{cl}+n_{3}^{qu}).
  \label{eq:g2n}
\end{align}
Taking into account the fact that 1-1 propagators vanish and the
parton states in the nuclear wave functions only contract with ``2"
fields, the number of 2-2 propagators is given by
\begin{align}\label{eq:n22def}
  n_{22}&=\frac{1}{2}(n_{2}^{ext}+\text{number of ``2" fields}
  -\text{number of ``1" fields}-2n_q) \nonumber\\
  &=\frac{1}{2}( n_{2}^{ext}-2 n_q + 2 n_{4}^{cl} - 2
  n_{4}^{qu}+n_{3}^{cl}-3n_{3}^{qu})
\end{align}
with $n_{2}^{ext}$ the number of external gluon ``2" fields in the
operator $O$, i.e., $G_{22}^{a\mu,b\nu}$. Plugging (\ref{eq:g2n}) into
(\ref{eq:n22def}) gives
\begin{align}\label{eq:pc}
  n_{22}&=\frac{1}{2}(n_{2}^{ext} + n_D)-n_q-2n^{qu}
\end{align}
with $n^{qu}=n_3^{qu}+n_4^{qu}$ the number of (quantum) vertices with
three ``1" fields.

The second ingredient of our proof is that the diagram gives a
non-vanishing contribution in the eikonal approximation only if there
is at least one $S^{(0)}_{22}$ on each quark line. Otherwise, its
contribution will be canceled by other diagrams. Let us only single
out one quark line without any $S^{(0)}_{22}$ propagator. Assume that
there are $n$ gluon lines connected to it. First, we take $n=2$. Let
us include the diagram with the two gluon lines connected in the
opposite order (while the rest part of the diagram is kept the
same). The two diagrams cancel with each other due to the following
cancellation
\begin{align}
\begin{array}{l}
\includegraphics[width=6cm]{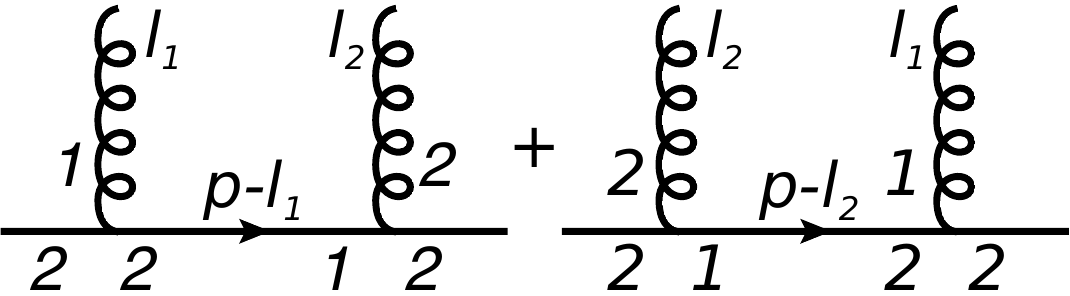}
\end{array}
\propto I_2\equiv \frac{1}{-l_1^\mp-i\epsilon}+\frac{1}{-l_2^\mp+i\epsilon}=0
\end{align}
with $l_1^\mp$ respectively corresponding to the diagrams with the
quark being from nucleus 1 or 2. Here, we have used the fact that
$l_1^\pm+l_2^\pm\approx 0$ in the eikonal limit.

The case with arbitrary $n$ attached gluon lines can be proved by
induction. Let us assume that the diagram with one quark line without
any $2-2$ propagators will be canceled by $n-1$ other diagrams. These
$n$ diagrams differ from each other only in the ways how the $n$ gluon
lines are connected to the quark line as shown in the following
diagram:
\begin{center}
  $
  \begin{array}{l}
    \includegraphics[width=0.8\textwidth]{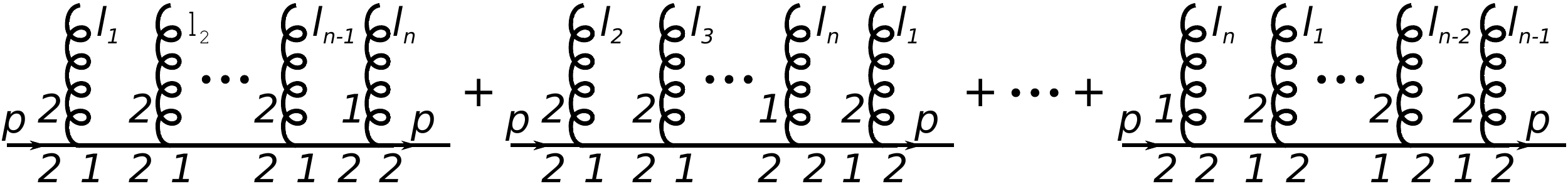}
  \end{array}.
  $
\end{center}
Since the rest of the diagrams are the same, they differ only in the
expressions from this quark line. Equivalently, we assume that the
above diagrams cancel with each other, that is
\begin{align}\label{eq:cancel}
  &I_n=\frac{1}{l_1(l_1+l_2)\cdots (l_1+\cdots+l_{n-1})}+\frac{1}{l_2(l_2+l_3)\cdots (l_2+\cdots+l_{n-1})(-l_1)}\nonumber\\
  &+\frac{1}{l_3(l_3+l_4)\cdots (l_3+\cdots+l_{n-1})(-l_1-l_2)(-l_2)}+\cdots\nonumber\\
  &+\frac{1}{(-l_1-l_2-\cdots-l_{n-1})\cdots
    (-l_{n-2}-l_{n-1})(-l_{n-1})}=0.
\end{align}
Here, each term corresponds to each diagram in the above figure. We
drop the superscript $\mp$ and the $i\epsilon$ prescription of all the
momenta $l_i$, which do not matter for our proof.

At the end we need only prove that (\ref{eq:cancel}) is also true for
$n+1$. Using (\ref{eq:cancel}) in the last term of $I_{n+1}$ we write
this last term as
\begin{align}\label{eq:In1}
  I_{n+1}^{(n+1)}&\equiv \frac{1}{(-l_1-\cdots-l_{n-1}-l_{n})\cdots
    (-l_{n-1}-l_{n})(-l_{n})}
  \nonumber\\
  &=\frac{1}{l_n}\left[ \frac{1}{l_1(l_1+l_2)\cdots (l_1+\cdots+\tilde l_{n-1})}+\frac{1}{l_2(l_2+l_3)\cdots (l_2+\cdots+\tilde l_{n-1})(-l_1)}\right.\nonumber\\
  &\qquad\left.+\cdots+\frac{1}{\tilde
      l_{n-1}(-l_1-l_2-\cdots-l_{n-2})\cdots
      (-l_{n-3}-l_{n-2})(-l_{n-2})}\right].
\end{align}
Then, by using the identity
\begin{align}
  \frac{1}{l_n(l+l_{n-1}+l_n)}=\frac{1}{l+l_{n-1}}\left(\frac{1}{l_n}-\frac{1}{l+l_{n-1}+l_n}\right),
\end{align}
we obtain
\begin{align}
I_{n+1}^{(n+1)}
=&-\frac{1}{l_1(l_1+l_2)\cdots (l_1+\cdots+l_{n-1}) (l_1+\cdots+l_{n})}-\frac{1}{l_2(l_2+l_3)\cdots (l_2+\cdots+l_{n})(-l_1)}\nonumber\\
&\qquad-\cdots-\frac{1}{ l_{n-1}(l_{n-1}+l_{n})(-l_1-\cdots-l_{n-2})\cdots (-l_{n-3}-l_{n-2})(-l_{n-2})}\nonumber\\
&\qquad-\frac{1}{ l_{n}(-l_1-\cdots-l_{n-1})\cdots (-l_{n-2}-l_{n-1})(-l_{n-1})}.
\end{align}
This exactly cancels the other terms in $I_{n+1}$. By induction,
(\ref{eq:cancel}) is true for all $n$.

Finally, by using the power counting (\ref{eq:pc}) and the identity
(\ref{eq:cancel}), we can make the following statements about the
classical field limit in our formalism:
\begin{enumerate}
\item{\it The diagrams for the two-point gluon correlator
    $G_{22}^{\mu\nu}$ should be of order $g^{n_D}$ with $n_D \ge 4 n_q
    -2$}. \\
  Indeed, \eq{eq:pc} with $n_2^{ext}=2$ gives
  \begin{align}
    \label{eq:pc2}
    n_D = 2 n_{22} + 2 n_q -2 + 4 n^{qu} \ge 4 n_q -2 + 4 n^{qu} \ge 4 n_q -2,
  \end{align}
  where we have used the fact that $n_{22} \ge n_q$ since, for the
  diagram not to cancel, each valence quark line should contain a 2-2
  propagator due to the proof above. The lowest possible value of
  $n_D$ corresponds to the classical dynamics. It is reached if
  $n_{22} = n_q$ and $n^{qu} =0$. The latter condition means no
  vertices with three ``1'' fields in the diagrams for the classical
  correlator. This is consistent with the conclusion in the functional
  approach \cite{Mueller:2002gd}. Each quark line has one
  $S^{(0)}_{22}$, which separates the diagram into two sub-diagrams
  connected by the cut quark propagators. {\it We conclude that
    classical $G_{22}^{\mu\nu}$ must be a product of two classical
    gluon fields.}

\item{\it $G_{12}^{\mu\nu}=0= G_{21}^{\mu\nu}$ at each order of
    $\frac{1}{g^2} (g^4 A^{\frac{1}{3}})^{n_q}$.} \\  Since
  now $n_2^{ext}=1$, \eq{eq:pc} gives
  \begin{align}
    \label{eq:pc3}
    n_D = 2 n_{22} + 2 n_q -1 + 4 n^{qu} \ge 4 n_q -1 + 4 n^{qu} \ge 4 n_q -1,
  \end{align}
  which is a higher order of the coupling than the classical $4 n_q
  -2$. Therefore, all order-$g^{4 n_q -2}$ diagrams should cancel.
\end{enumerate}


\section{The Boltzmann equation for gluons}
\label{B}

In this Appendix we review the standard derivation of Boltzmann
equation for gluons.  Let us define
\begin{align}
  \square^{\mu\nu}_x\equiv
  g^{\mu\nu}\square_x-\partial_x^\mu \partial_x^\nu-\frac{n^\mu
    n^\nu}{\xi}\qquad\text{with $\xi\to 0$}.
\end{align}
From the Dyson-Schwinger equation for gluons, one can get
\begin{align}
  {\square^{\mu}_x}_\rho G^{a\rho, b\nu}_{22}(x,y)=&\int d^4 z~{\Pi_{11}^{a\mu}}_{c\rho}(x,X+z)G^{c\rho, b\nu}_{12}(X+z,y)\nonumber\\
  &+\int d^4 z~{\Pi_{12}^{a\mu}}_{c\rho}(x,X+z)G^{c\rho, b\nu}_{22}(X+z,y),\\
  {\square^{\mu}_y}_\rho G^{ b\nu,a\rho}_{22}(x,y)=&\int d^4 z~G^{ b\nu,c\rho}_{21}(x, X+z) {{\Pi_{11}}_{c\rho}}^{a\mu}(X+z,y)\nonumber\\
  &+\int d^4 z~G^{ b\nu,c\rho}_{22}(x, X+z)
  {{\Pi_{21}}_{c\rho}}^{a\mu}(X+z,y),\label{eq:Gbnuarho}
\end{align}
with $X^\mu\equiv \frac{x^\mu+y^\nu}{2}$ and $\Pi$'s being
self-energies. Accordingly,
\begin{align}
  \square^{\mu}_{\rho} G^{a\rho, b\nu}_{22}(X,p)=&\int d^4 \Delta x e^{ip\cdot \Delta x} d^4 z\int \frac{d^4 p'}{(2\pi)^4}\frac{d^4 p''}{(2\pi)^4} e^{-i p'\cdot\left(\frac{\Delta x}{2}-z\right)-i p''\cdot\left(\frac{\Delta x}{2}+z\right)}\nonumber\\
  &\times\left[{\Pi_{11}^{a\mu}}_{c\rho}\left(X+\frac{z+\frac{\Delta x}{2}}{2},p'\right)G^{c\rho, b\nu}_{12}\left(X+\frac{z-\frac{\Delta x}{2}}{2},p''\right)\right.\nonumber\\
  &\left.+{\Pi_{12}^{a\mu}}_{c\rho}\left(X+\frac{z+\frac{\Delta x}{2}}{2},p'\right)G^{c\rho,
      b\nu}_{22}\left(X+\frac{z-\frac{\Delta x}{2}}{2},p''\right)\right],
\end{align}
where $\Delta x^\mu\equiv x^\mu-y^\mu$, and
\begin{align}
  \square^{\mu}_{\rho}\equiv
  \left[g^\mu_\rho\left(\frac{1}{2}\partial_X-ip\right)^2-\left(\frac{1}{2}\partial_X-ip\right)^\mu\left(\frac{1}{2}\partial_X-ip\right)_\rho-\frac{n^\mu
      n_\rho}{ \xi}\right].
\end{align}
By assuming that $x^\mu$ and $z^\mu$ are negligible compared to
$X^\mu$ in $G^{a\mu,b\nu}$ and $\Pi^{a\mu,b\nu}$, one has
\begin{align}
  \square^{\mu}_{\rho}G^{a\rho, b\nu}_{22}(X,p)=
  {\Pi_{11}^{a\mu}}_{c\rho}\left(X,p\right)G^{c\rho,
    b\nu}_{12}\left(X,p\right)+{\Pi_{12}^{a\mu}}_{c\rho}\left(X,p\right)G^{c\rho,
    b\nu}_{22}\left(X,p\right).\label{eq:ds01}
\end{align}
In this approximation, (\ref{eq:Gbnuarho}) gives
\begin{align}
  \square^{*\mu}_{\rho}G^{b\nu, a\rho}_{22}(X,p)=G^{
    b\nu,c\rho}_{21}(X,p) {{\Pi_{11}}_{c\rho}}^{a\mu}(X,p)+G^{
    b\nu,c\rho}_{22}(X, p)
  {{\Pi_{21}}_{c\rho}}^{a\mu}(X,p).\label{eq:ds02}
\end{align}
By symmetry, one has
\begin{align}
  G^{ b\nu,c\rho}_{22}(X,p)=G^{c\rho, b\nu}_{22}(X,p),\qquad
  {{\Pi_{11}}_{c\rho}}^{a\mu}(X,p)={{\Pi_{11}}^{a\mu}}_{c\rho}(X,p).
\end{align}
Subtracting (\ref{eq:ds02}) from (\ref{eq:ds01}) gives
\begin{align}
  &\left[-2i g^\mu_\rho p\cdot \partial_X+ip^\mu\partial_{x^\rho}+ip_\rho \partial_X^\mu\right]G^{a\rho, b\nu}_{22}\nonumber\\
  &\qquad=\left[{\Pi_{12}^{a\mu}}_{c\rho}-{{\Pi_{21}}_{c\rho}}^{a\mu}\right]G^{c\rho,
    b\nu}_{22}+{\Pi_{11}^{a\mu}}_{c\rho}\left[G^{c\rho, b\nu}_{12}-G^{
      b\nu,c\rho}_{21}\right]. \label{B9}
\end{align}

By using the ansatz \cite{Kadanoff,Chou:1984es,Blaizot:2001nr} 
\begin{align}
  G_{21}^{a\mu,b\nu}(X,p)&=\delta^{ab}\left(-g^{\mu\nu}+\frac{p^\mu n^\nu+p^\nu n^\mu}{n\cdot p}\right)G_R(p),\nonumber\\
  G_{12}^{a\mu,b\nu}(X,p)&=\delta^{ab}\left(-g^{\mu\nu}+\frac{p^\mu n^\nu+p^\nu n^\mu}{n\cdot p}\right)G_A(p),\nonumber\\
  G_{22}^{a\mu,b\nu}(X,p)&=2\pi\delta^{ab}\left(-g^{\mu\nu}+\frac{p^\mu
      n^\nu+p^\nu n^\mu}{n\cdot
      p}\right)\left[f(X,p)+\frac{1}{2}\right]\delta(p^2),\label{eq:quasipart}
\end{align}
and contracting \eq{B9} with
$\sum_\lambda\epsilon_{\lambda\nu}(p)\epsilon^*_{\lambda\mu}(p)\delta^{ab}$,
one has
\begin{align}
  p\cdot \partial f = &\frac{i\delta^{ab}}{ 4(N_c^2-1) }\sum_\lambda\epsilon_{\lambda\nu}(p)\epsilon^*_{\lambda\mu}(p)\left[\Pi_{21}^{b\nu,a\mu}-\Pi_{12}^{a\mu,b\nu}\right]\left(f+\frac{1}{2}\right)\nonumber\\
  &+\frac{i\delta^{ab}}{ 4(N_c^2-1)
  }\text{Sign}(p^0)\sum_\lambda\epsilon_{\lambda\nu}(p)\epsilon^*_{\lambda\mu}(p)\Pi_{11}^{a\mu,b\nu}\nonumber\\
  =&\frac{i}{ 2 }\left[\bar\Pi_{21}-\bar\Pi_{12}\right]\left(f+\frac{1}{2}\right)+\frac{i}{ 2
  }\text{Sign}(p^0)\bar\Pi_{11}
\end{align}
with $f$ the distribution function. Since $G_{22}^{a\mu,b\nu}(x,y)$ is
real, it satisfies
\begin{align}
G_{22}^{a\mu,b\nu}(X,p)=G_{22}^{a\mu,b\nu}(X,-p).
\end{align}
Hence, one only needs to solve for $f$ at positive $p^0$, which
satisfies
\begin{align}
  p\cdot \partial f = &\frac{i}{ 2 }\left[\bar\Pi_{21}-\bar\Pi_{12}\right]\left(f+\frac{1}{2}\right)+\frac{i}{ 2
  }\bar\Pi_{11}.\label{eq:gluonBolt}
\end{align}

Inserting (\ref{eq:Pi11}) and (\ref{eq:Pi21}) into the above equation
and ignoring $\frac{1}{2}$ associated with $f$ in (\ref{eq:quasipart})
and the above equation gives the Boltzmann equation in the classical
limit (see \cite{Mathieu:2014aba} for another derivation). If one
keeps $\frac{1}{2}$'s \cite{Mueller:2002gd}, one has
\begin{align}
  p\cdot \partial f=\frac{1}{4}\int_{p_1,p_2,p_3} &(2\pi)^4\delta(p+p_1-p_2-p_3)\overline{|M|^2}[f_2f_3(f+1)(f_1+1)\nonumber\\
  &-ff_1(f_2+1)(f_3+1)+\frac{1}{4}(f_2+f_3-f_1-f)].\label{eq:boltEqua}
\end{align}
Just like in the $\lambda\varphi^4$ theory \cite{Epelbaum:2014mfa},
the last term on the right-hand side of (\ref{eq:boltEqua}) leads to a UV
divergence. This qualitatively helps understand the origin for the
lattice spacing dependence observed in \cite{Gelis:2013rba} although a
quantitative analysis requires the calculation in lattice QCD (see a
discussion in QED \cite{Epelbaum:2015cca,Epelbaum:2015vaa}).

\begin{figure}
\begin{center}
\includegraphics[width=0.7\textwidth]{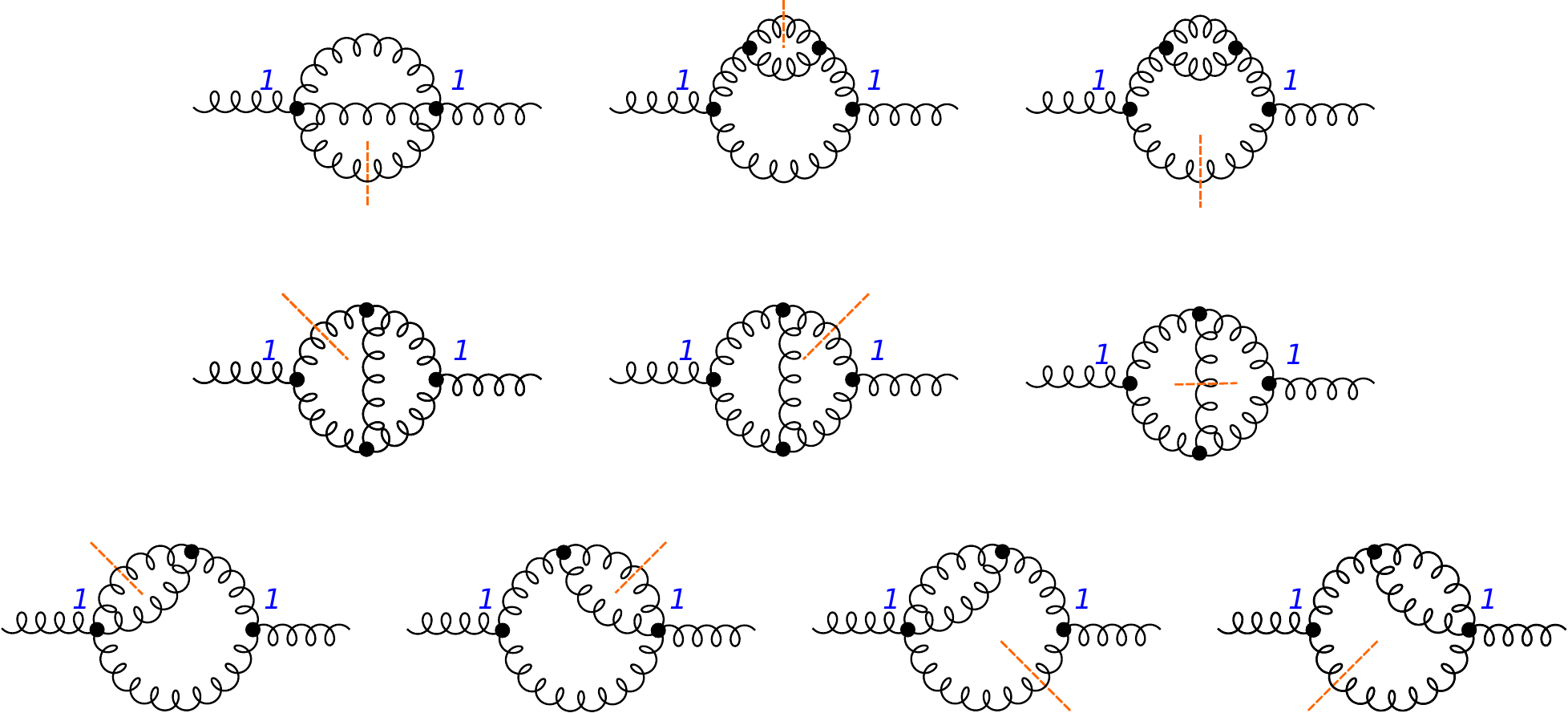}
\end{center}
\caption{Diagrams for $\Pi_{11}^{a\mu,b\nu}$.}
\label{fig:Pi11q}
\end{figure}

\begin{figure}
\begin{center}
\includegraphics[width=\textwidth]{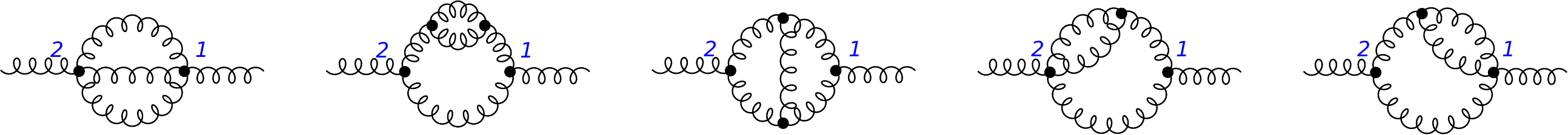}
\end{center}
\caption{Diagrams for $\Pi_{21}^{a\mu,b\nu}$.}
\label{fig:Pi21q}
\end{figure}

In order to cancel the above UV divergence, one needs to include the
contributions from the diagrams in Figs. \ref{fig:Pi11q} and
\ref{fig:Pi21q}. The diagrams in Fig. \ref{fig:Pi11q} give
\begin{align}
  -i\bar\Pi_{11}(X,p)=&-\frac{1}{8}\int_{p_1,p_2,p_3} (2\pi)^4\delta(p+p_1-p_2-p_3)\overline{|M|^2} \nonumber\\
  &\times[ g_{22}(X,p_1)-g_{22}(X,p_2)-g_{22}(X,p_3)].\label{eq:Pi11q}
\end{align}
And the diagrams in Fig. \ref{fig:Pi21q} yield
\begin{align}
  -i&[\bar\Pi_{21}(X,p)-\bar\Pi_{12}(X,p)]=-\frac{1}{8}\int_{p_1,p_2,p_3} (2\pi)^4\delta(p+p_1-p_2-p_3)\overline{|M|^2}.\label{eq:Pi21q}
\end{align}

Plugging (\ref{eq:Pi11}), (\ref{eq:Pi21}), (\ref{eq:Pi11q}) and
(\ref{eq:Pi21q}) into the right-hand side of
(\ref{eq:gluonBolt}) gives the Boltzmann equation for gluons
\begin{align}
  p\cdot \partial f =&\frac{1}{4}\int_{p_1,p_2,p_3} (2\pi)^4\delta(p+p_1-p_2-p_3)\overline{|M|^2}\nonumber\\
  &\times[f_2f_3(f+1)(f_1+1)-ff_1(f_2+1)(f_3+1)]
\end{align}
with $\overline{|M|^2}$ given in (\ref{eq:Msgggg}).




\providecommand{\href}[2]{#2}\begingroup\raggedright\endgroup
  

\end{document}